\documentclass[1p]{elsarticle}

\usepackage{hyperref,tabularx,amsmath,amssymb,bm, mathtools}
\usepackage[list=true, font=normalsize, labelfont=sf,
labelformat=brace, position=top]{subcaption}
\usepackage{pgfplots}
\usepackage{tikz}
\usepackage[c1]{optidef}

\usepackage{xspace}

\journal{Tribology International}

\begin{document}

\begin{frontmatter}

\title{Modeling Ice Friction for Vehicle Dynamics of a Bobsled with Application in Driver Evaluation and Driving Simulation\tnoteref{mytitlenote}}
\tnotetext[mytitlenote]{This work was generously supported by BMW AG.}

\author{Julian von Schleinitz\fnref{bmw} \corref{mycorrespondingauthor}}
\author{Lukas Wörle \fnref{bmw}}
\author{Michael Graf \fnref{ge}}
\author{Andreas Schröder \fnref{us}}

\fntext[bmw]{BMW AG}
\fntext[ge]{Graf Engineering}
\fntext[us]{University of Salzburg}


\cortext[mycorrespondingauthor]{Corresponding author}

\begin{abstract}

We provide an ice friction model for vehicle dynamics of a two-man bobsled which can be used for driver evaluation and in a driver-in-the-loop simulator. Longitudinal friction is modeled by combining experimental results with finite element simulations to yield a correlation between contact pressure and friction. 
To model lateral friction, we collect data from 44 bobsleigh runs using special sensors. Non-linear regression is used to fit a bob-specific one-track vehicle dynamics model to the data. It is applied in driving simulation and enables a novel method for bob driver evaluation. Bob drivers with various levels of experience are investigated. It shows that a similar performance of the top drivers results from different driving styles.

\end{abstract}

\begin{keyword}
Ice Friction, Bobsleigh,  Driver Evaluation, Vehicle Dynamics
\end{keyword}

\end{frontmatter}

\section{Introduction}
In the sport of bobsleigh, the driver, the bobsled and especially the interaction between both are a crucial part of the overall performance. A key element in understanding this system is the friction between the runners and the ice surface. The friction of steel on ice is very low, enabling the possibility of fast wintersports like luge, bobsleigh or skeleton  reaching top speeds of over $150$kph. However, the ultra low friction of ice arises from complex material behavior and is a matter of current research \cite{kietzig_ice_2010}. Therefore, specific experiments and simulations are  important  to make further advances in this science area and to enable performance improvements in winter sports.\\ 
Most experiments are carried out under simplified conditions (see Section \ref{chap_related}) compared to a highly dynamic bob run on a track. In particular, the lateral friction of a bob runner, which occurs when the velocity vector of the runner does not align with its longitudinal axis, is often not considered. This is surprising given that lateral friction is approximately one order of magnitude higher than longitudinal friction and has significant implications on  athlete and bobsled performance. In general, vehicle dynamics in combination with ice friction is sparsely researched. This work aims to close this gap by providing a friction model for a bobsled with application in the area of driver evaluation and driving simulation. \\
The friction model is created and validated with real-world measurement data. In the case of lateral friction, which is crucial for realistic driving behavior in a simulator, the data originates from measurements with a professional bobsled suited for international competitions under race conditions. Our generated friction model is implemented at the BMW bobsled simulator  which is being used by the German bobsleigh national team for preparation for the 2022 Olympic Winter Games.

\subsection{Related work}
\label{chap_related}
\subsubsection{Longitudinal friction}

The majority of research for steel-ice friction focuses on longitudinal friction, i.e.~ the frictional force which acts against the driving direction when driving straightly. 
There are different kinds of experimental setups to investigate ice friction.  For example, \citet{hainzlmaier_tribologically_2005} utilized an iced centrifuge with a slider gliding over a flat surface.
\citet{Scherge.2013} used a modified tire test bench, where the slider glides in a concave curvature  and determined a range for the longitudinal friction coefficient $\mu_\mathrm{x} = 7-16 \cdot 10^{-3}$. \citet{Scherge.2018} also tested the influence of speed and temperature and found a lower limit for the friction coefficient  $\mu_\mathrm{x} \geq 8 \cdot 10^{-3}$ under their experimental conditions.  Unlike rotational devices, linear devices are better suited to investigate ice friction on a fresh surface \cite{kietzig_ice_2010}. \citet{Makkonen.2014} also pointed out that devices where a slider repeatedly glides over the same surface could lead to misinterpretation due to frictional heating. \citet{hasler_novel_2016} conducted  ski-sport specific tests on a $24m$ linear tribometer. It was also used to optimize luge steels for the Austrian luge national team.\\
 Real-life experiments are closely correlated to the application area. \citet{dekoning_ice_1992}  provided friction coefficients for speed skating by utilizing strain gauges ($\mu_\mathrm{x}=4.6 - 5.9 \cdot 10^{-3}$). \citet{poirier_dis} investigated ice friction on a bobsleigh track during a bob race with a radar gun ($\mu_\mathrm{x}=3.6 - 5.3 \cdot 10^{-3}$). They performed friction experiments  in an ice house as well ($\mu_\mathrm{x}=4.2 \cdot 10^{-3}$) \cite{Poirier.2011}. \citet{Irbe.2021} performed tests with a skeleton  at the start ramp of a track ($\mu_\mathrm{x}=4.3 - 6.72 \cdot 10^{-3}$). \citet{Lozowski.2013} developed a numerical model for longitudinal friction of a bobsled runner and compared different runner geometries on a flat surface and determined $\mu_\mathrm{x} \approx 4.6-4.8  \cdot 10^{-3}$ for rocker radii (i.e. radii of the runners in longitudinal direction) between 20 and 50m and cross section radii from 4-7mm. All these real-life experiments were only carried out  on straight sections of a track. An exception is  \citet{Mossner.2011} who  determined $\mu_\mathrm{x}=10 - 12 \cdot 10^{-3}$ for luge on the track in Vancouver, which is substantially higher than the other real life experiments. However it is still in the range of some laboratory experiments, e.g. \cite{Scherge.2013}.\\
While the above mentioned studies show significant deviations and were not executed under the same conditions, most real life experiments for bobsleigh are  located at  around $\mu_\mathrm{x} \approx 4  \cdot 10^{-3}$. \\
An important factor of influence on ice friction is the contact pressure which is exerted on the ice by the gliding material \cite{hainzlmaier_tribologically_2005}. \citet{hainzlmaier_tribologically_2005} reported a pressure dependency of the longitudinal friction coefficient. \citet{Liefferink.2021} also stated that pressure, temperature and speed have an important impact on ice friction, which is confirmed by \citet{Velkavrh.2019}.  Unlike ice temperature, gliding speed or normal force, the pressure which is exerted on the ice can be influenced by altering the contact area of the  bobsled runners, making it a very important parameter in practice. This is supported by the considerable effort that bob teams around the world put into optimizing the shape of  bobsled runners.

\subsubsection{Lateral friction}
Literature investigating the lateral friction of a bobsled is sparser even though it is very important for its driving behavior. Most studies were conducted in the context of a bob simulator or simulation model. \citet{Rempfler.2015} developed a bob simulator and found a lateral friction coefficient of $\mu_\mathrm{y}= 0.02- 0.1$ to reproduce realistic driving behavior in the simulator, which is also described in \cite{Rempfler.2016}.
\citet{arnold_analyse_2013} calculated an upper bound for the  lateral friction coefficient based on results provided by \citet{hainzlmaier_tribologically_2005} with  $\mu_\mathrm{y} \leq 0.11- 0.28$. According to \citet{Scherge.2021}, the lateral friction for a bobsled is ten times higher than longitudinal friction, which would account for $\mu_\mathrm{y}= 0.05- 0.09$.
\citet{Braghin.2011} developed a bob driver model for bobsled optimization and provided  an equation for the lateral friction force $F_\mathrm{y}$ based on the normal force $F_\mathrm{z}$ and the side slip angle of the runner $\alpha$
\begin{equation}
\label{eq_fric_bra}
F_\mathrm{y} = \mu_\mathrm{y} F_\mathrm{z} \frac{2}{\pi} \mathrm{atan}(k_3 \alpha),
\end{equation}
 whereby it holds for the parameter $k_3= 50 \mathrm{rad}^{-1}$ and $\mu_\mathrm{y} = 0.5$ \cite{Braghin.2011b}. It seems that they used this equation for both front and rear axle. 
 To sum up, the friction coefficients reported in literature seem to be highly dependent on the experimental circumstances, as the deviations are quite high. We believe that a certain amount of the deviations can be explained by a combination of lateral and longitudinal friction. Especially, when determining friction on a track where lateral friction is needed to achieve the desired trajectory, combined friction needs to be accounted for, e.g. by using a dynamic vehicle model. This may be an explanation as to why the friction coefficient reported by \citet{Mossner.2011} is so high: it is an average over a complete track and therefore also includes  the contribution of lateral friction.  \\
 
 \subsubsection{Driver evaluation}
The research into bob driver evaluation is a field with very few publications. This is interesting since there is high potential for performance improvements for both athlete and bobsled development. In contrast, in motorsport considerable effort is put into adjusting the setup of a race car to the driver's driving style \cite{Schleinitz.2019,Worle.20,Schleinitz.2021b}, with driving simulators  also being used for that task \cite{Schwarzhuber.2020,Schwarzhuber.2020b}. \citet{Schleinitz.2021} presented a method for race driver evaluation using a tire grip potential exploitation rating, i.e. it was  analyzed how much of the available grip was utilized by the driver. The method we present is similar in some sense. Since bob drivers cannot actively accelerate in the driving direction, the best they can do is to minimize the frictional losses. \\
Though, not directly intended for driver evaluation, there are some works  dealing with simulation models which could be used for that task. For example \citet{Gong.2016} created a simple steering model for a skeleton simulation and analyzed different control strategies. \citet{Zhang.1995} developed steering models for bobsleigh using optimal control. However, for both cases, a virtual model of a track has to be created, therefore it is not possible to directly evaluate drivers using only real world data. \\

 \subsection{Overview}
 
 \begin{figure}
\centering
\begin{subfigure}[c]{0.99\textwidth}
\includegraphics[width=0.99\textwidth]{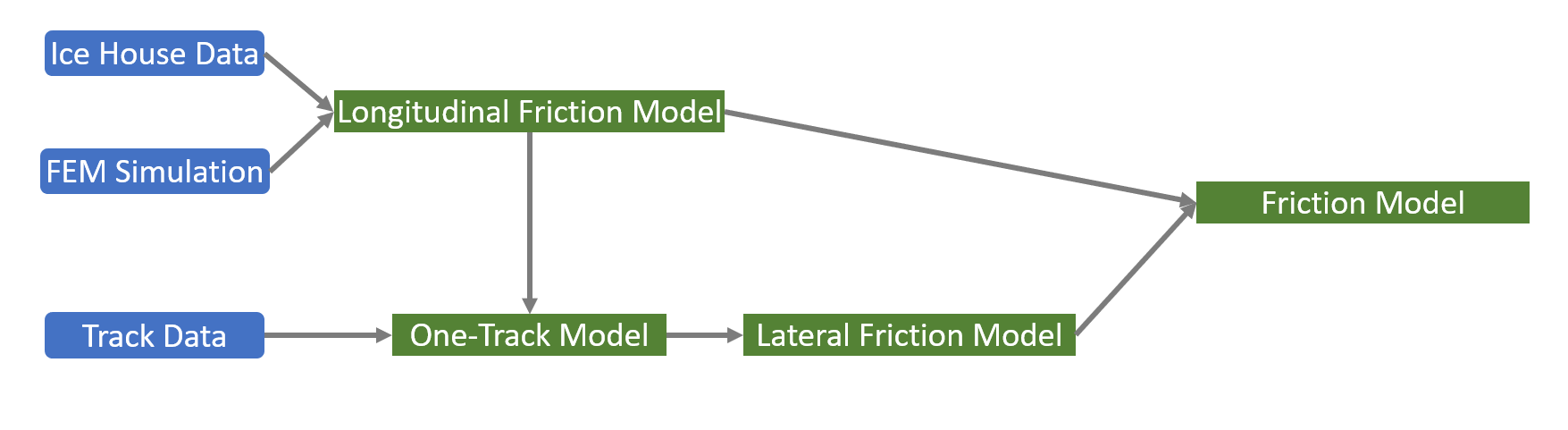}
\subcaption{Model generation: Ice house experiments, FEM simulations and track data are combined to a longitudinal and lateral friction model.}
\end{subfigure}
\begin{subfigure}[c]{0.99\textwidth}
\includegraphics[width=0.99\textwidth]{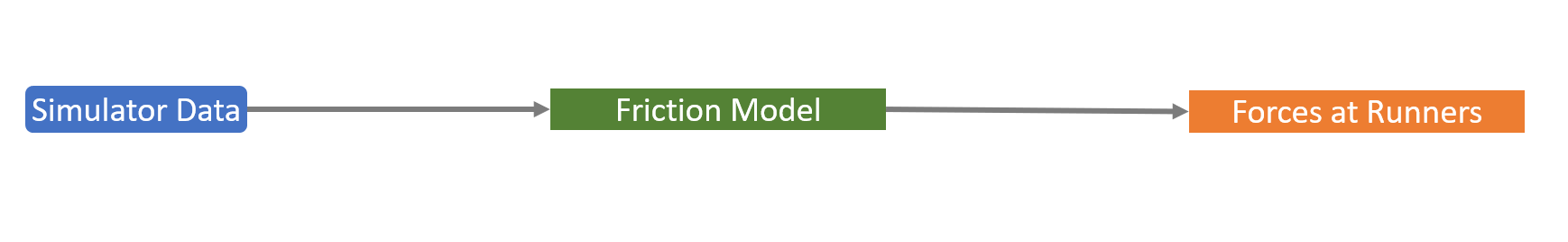}
\subcaption{Driving simulation: The friction model is applied to generate the resulting forces at the runners during the simulation.}
\end{subfigure}
\begin{subfigure}[c]{0.99\textwidth}
\includegraphics[width=0.99\textwidth]{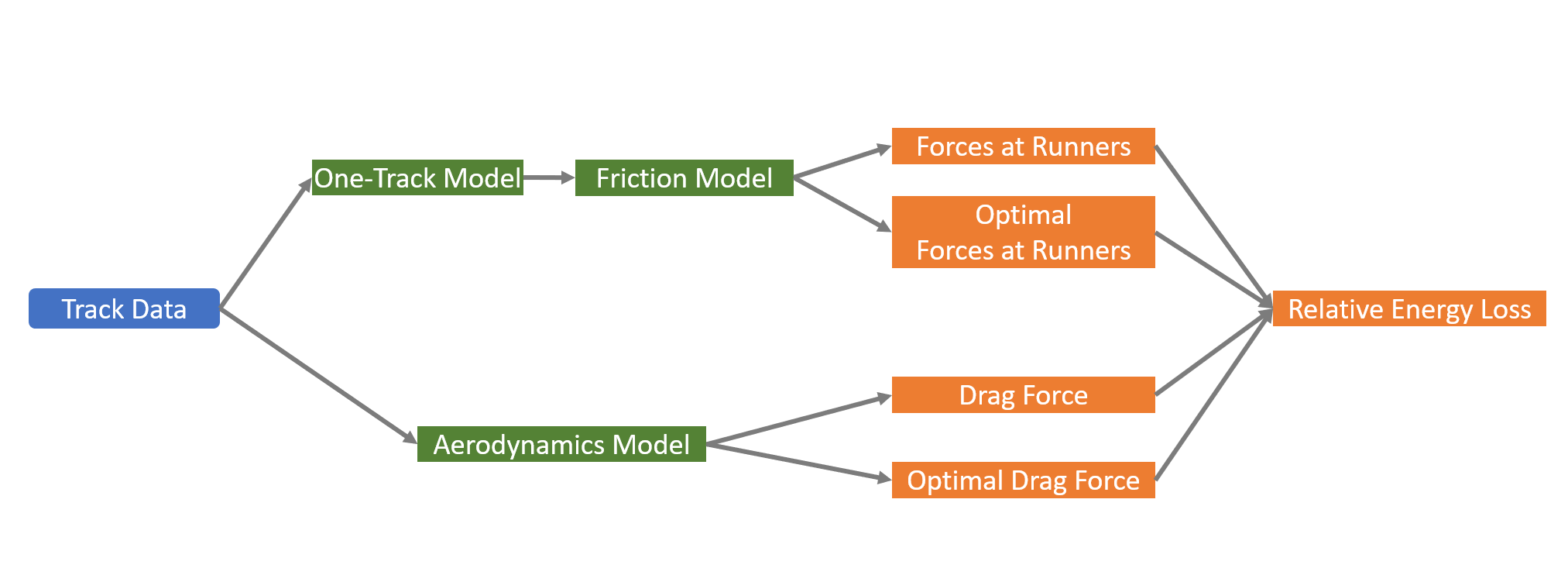}
\subcaption{Driver evaluation: The friction model is applied with inputs from the one-track model to yield the actual and optimal forces at the runners. In combination with the aerodynamic drag force a relative energy loss is determined. Using this definition, the smaller the energy loss the better the driving style is. }
\end{subfigure}
\caption{Graphical overview of this work.  }
\label{fig_overview}
\end{figure}

The rest of this work is structured as follows. Section 2 details the materials and methods. The aim is develop a model for both longitudinal and lateral friction. First, the data acquisition hardware is described. A friction experiment in an ice house is carried out to determine the longitudinal friction coefficient for various samples. We use a finite element contact simulation to obtain the contact pressure which is exerted on the ice. Both results are combined to an equation for the longitudinal friction force. Then, we create a bob-specific one-track model to describe its vehicle dynamics. Using data measured on a bob track, non-linear regression is utilized to fit a lateral friction model for the front and rear runner. The application of the friction model for a driving simulator is briefly sketched. 
What is more, the model is  used to evaluate bob drivers by comparing the calculated energy losses caused by ice friction and also aerodynamic drag. The key element is that both losses depend on the angle of the runner or the bob itself to the driving direction and can therefore be influenced by the driver. \\
Section 3 presents the results of this study. Parameters for the longitudinal and lateral friction model are shown and the vehicle dynamics model is validated and compared to another model from literature. Lastly, the driver evaluation method is applied to analyze five bob drivers with various levels of experience.\\

Summarized, the main contributions of this work are: (1) A bob-suitable vehicle dynamics model  for driving simulation which is real-time executable and can be used for analyzing ice friction. (2) A novel method for bob driver evaluation using real world measurement data.
(3) Application of the driver evaluation method  on a data set containing various drivers showing that different driving styles results in a similar overall performance.

\section{Materials and methods}

\subsection{Data acquisition}
In this paper we used measurement devices from automotive applications which are common in motorsport. \autoref{tab_sensors} lists the sensors and the measured quantities. Special equipment such as an optical speed-over ground sensor is needed to determine precisely the longitudinal and lateral velocity of the sled. For example the sensors  that are used in  Bob races and which are regulated by the international bobsleigh and skeleton federation (IBSF) are not sufficient for that purpose. The sample rate was between $100$ and $500$Hz depending on the sensor. The data was filtered using a low-pass filter and afterwards downsampled to  $100$Hz for further analysis.   The symbols and subscripts used in this paper are listed in and \autoref{tab:subscripts} and \autoref{tab:symbols}, respectively.  

\begin{table}[h]
	\centering 
	\caption{Description of the sub- and superscripts used in this paper.}
	\small
	\begin{tabular}{l l}
		\hline	
		 \multicolumn{1}{c}{\bfseries Subscripts} & \multicolumn{1}{c}{\bfseries Description} \\ \hline  
		 $\mathrm{(x,y,z)}$ &body coordinate system\\ \hline
		  $\mathrm{(\tilde{x},\tilde{y},\tilde{z})}$ & coordinate system aligned with driving direction\\ \hline
		 $\mathrm{f}$ &at front runner\\ \hline
		  $\mathrm{f0}$ &at front runner, unrotated\\ \hline
		   $\mathrm{r}$ &at rear runner\\ \hline
		   $\mathrm{cog}$ &at center of gravity\\ \hline
		    $\mathrm{ext}$ & external\\ \hline
		     $\mathrm{aero}$ & aerodynamic\\ \hline
		      $\mathrm{ice}$ & ice frictional\\ \hline
		        $\mathrm{ot}$ & one-track model\\ \hline
		           $\mathrm{fm}$ & friction model\\ \hline
		            $\mathrm{kin}$ & kinetic\\ \hline
		             $\mathrm{pot}$ & potential\\ \hline
		              $\mathrm{loss}$ & loss\\ \hline

	\end{tabular}
	\label{tab:subscripts}
\end{table}

\begin{table}
	\centering 
	\caption{Description of the symbols used in this paper.}
	\small
	\begin{tabular}{l l}
		\hline	
		 \multicolumn{1}{c}{\bfseries Symbols} & \multicolumn{1}{c}{\bfseries Description} \\ \hline
		 $v$ &  velocity \\ \hline
		$m$ & mass\\ \hline
				    $E$ & energy \\ \hline   
				    $s$ & distance \\ \hline  
				     $h$ & height \\ \hline  
				     $p$ & pressure \\ \hline  
				      $r$ & radius \\ \hline  
				       $\kappa$ & hill slope angle \\ \hline  
				      
		$F_\mathrm{x},~F_\mathrm{y},~F_\mathrm{z} $ & forces in x, y and z  \\ \hline
		$a_\mathrm{x},~a_\mathrm{y},~a_\mathrm{z} $ & accelerations in x, y and z  \\ \hline
		$M_\mathrm{x},~M_\mathrm{y},~M_\mathrm{z} $ & torque in x, y and z  \\ \hline
		$\varphi,~\dot{\varphi},~ \ddot{\varphi}$ & angle/ angular velocity/ acceleration around the roll axis (x)\\ \hline
		$\theta,~\dot{\theta},~ \ddot{\theta}$ & angle/ angular velocity/ acceleration around the pitch axis (y)\\ \hline
		$\psi, ~\dot{\psi},~ \ddot{\psi}$ & angle/ angular velocity/ acceleration around the yaw axis (z)\\ \hline
		$J_\mathrm{yy},~ J_\mathrm{zz}$ & principal moment of inertia around the y-/ z-axis\\ \hline
		 $\delta$ & steering angle \\ \hline
		 $\gamma$ & roll-split angle \\ \hline
		  $\alpha_\mathrm{f},\alpha_\mathrm{r}$ & side slip angles at the runners \\ \hline
		   $\beta$ & chassis side slip angle (at cog) \\ \hline
		 $l_\mathrm{F},~l_\mathrm{R}$ & distance front/rear axle to cog \\ \hline
		 $C_\mathrm{x}A_\mathrm{x}, ~C_\mathrm{y}A_\mathrm{y}$ & aerodynamic drag area \\ \hline
				 $R,~T$ & air parameters: ideal gas constant, temperature \\ \hline 
		$\mu_\mathrm{x},~\mu_\mathrm{y}$ & friction coefficient in x-/y-direction\\ \hline
		 $C_\mathrm{y},~E_\mathrm{y},~K_\mathrm{y},~\zeta_\mathrm{y}$ & friction model parameters in y-direction \\ \hline
		 		 $B_\mathrm{x},~C_\mathrm{x},~D_\mathrm{x},~E_\mathrm{x}, ~\zeta_\mathrm{x}$ & friction model parameters in x-direction \\ \hline
		  $\bm{A}$ &rotation matrix \\ \hline
		\end{tabular}
	\label{tab:symbols}
\end{table}

\begin{table}[bt]
\caption{Utilized Sensors and measured parameters}
\label{tab_sensors}
\begin{tabularx}{\textwidth}{XX}
\hline
Sensor&Measured parameter \\
\hline
Accelerometer&Accelerations: $a_\mathrm{x},~a_\mathrm{y},~a_\mathrm{z} $\\
Gyroscope & Rotational velocities: $\dot{\varphi},~\dot{\theta},~\dot{\psi} $\\
Optical Speed-over Ground Sensor & Velocity in the xy-plane: $v$\\
&Side slip angle: $\alpha$\\
Rotational Potentiometer&  Steering angle: $\delta$ \\
&Roll-split angle: $\gamma$ \\
\hline
\end{tabularx}
\end{table}



\subsection{Longitudinal friction model}
\label{sec:fric_long}
Measuring the ultra-low longitudinal frictional forces during a bobsleigh run is very challenging because of vibrations and much higher forces orthogonal to the driving direction. Therefore we carry out experiments in a controlled environment. This section is based on preliminary studies (mainly the unpublished master thesis `Analysis and Simulation of Ice Friction in the Wintersport of Luge' by J.v. Schleinitz). 
Previous studies showed that ice friction is dependent on the pressure which is exerted by the gliding body \cite{hainzlmaier_tribologically_2005}. Therefore,  we conduct an experiment  in an ice house with a luge sled. The gliding experiments are carried out on an ice rink, i.e. an almost flat surface, the mass is kept constant and the contact area is changed. Luge steels have a sharp edge at the cross section, while bobsled runners are round. For our experiment, using luge steels has the advantage that a wide range of contact areas and therefore pressures can be studied. A finite element simulation is then used to determine the pressure which is exerted on the ice. As an approximation for the friction of bobsled runners, the resulting correlation between contact pressure and friction can be applied by simulating bobsled runner geometries to determine the contact pressure.  
 
\subsubsection{Ice house experiments}
\begin{figure}
    \centering
    \begin{subfigure}[b]{0.49\textwidth}
        \centering
        \includegraphics[width=\textwidth]{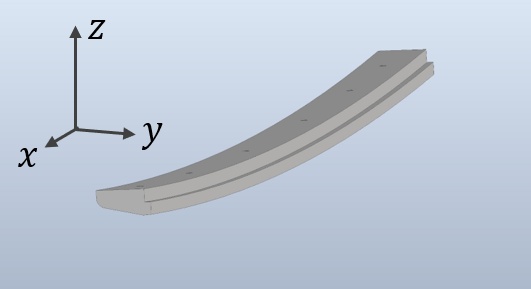}
        \caption{Isometric view}
    \end{subfigure}
    \hfill
    \begin{subfigure}[b]{0.49\textwidth}
        \centering
        \includegraphics[width=\textwidth]{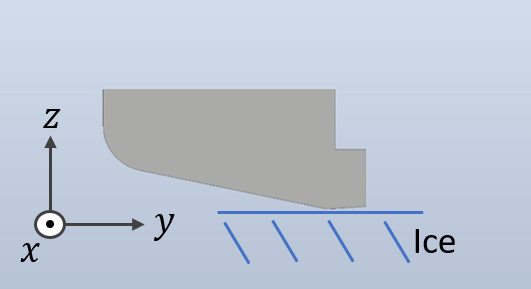}
        \caption{Cross section view}
    \end{subfigure}
        \begin{subfigure}[b]{0.49\textwidth}
        \centering
        \includegraphics[width=\textwidth]{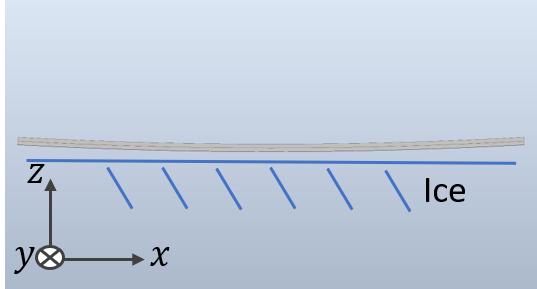}
        \caption{Longitudinal section view}
    \end{subfigure}
    \hfill
    \caption{Schematics of the shape of a luge steel.}
    \label{fig:luge_steel_shape}
\end{figure}

\begin{figure}
    \centering
    \begin{subfigure}[b]{0.49\textwidth}
        \centering
        \includegraphics[width=\textwidth]{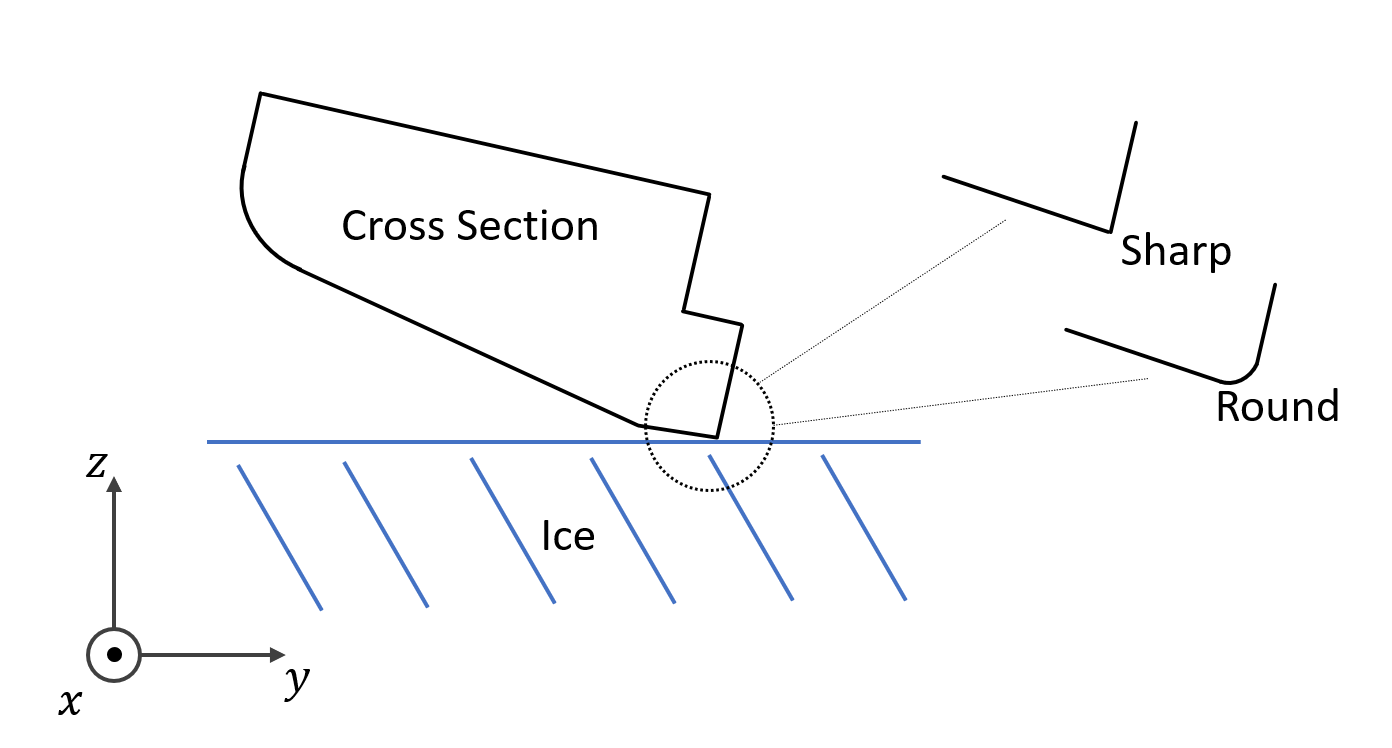}
        \caption{Cross section}
    \end{subfigure}
    \hfill
    \begin{subfigure}[b]{0.49\textwidth}
        \centering
        \includegraphics[width=\textwidth]{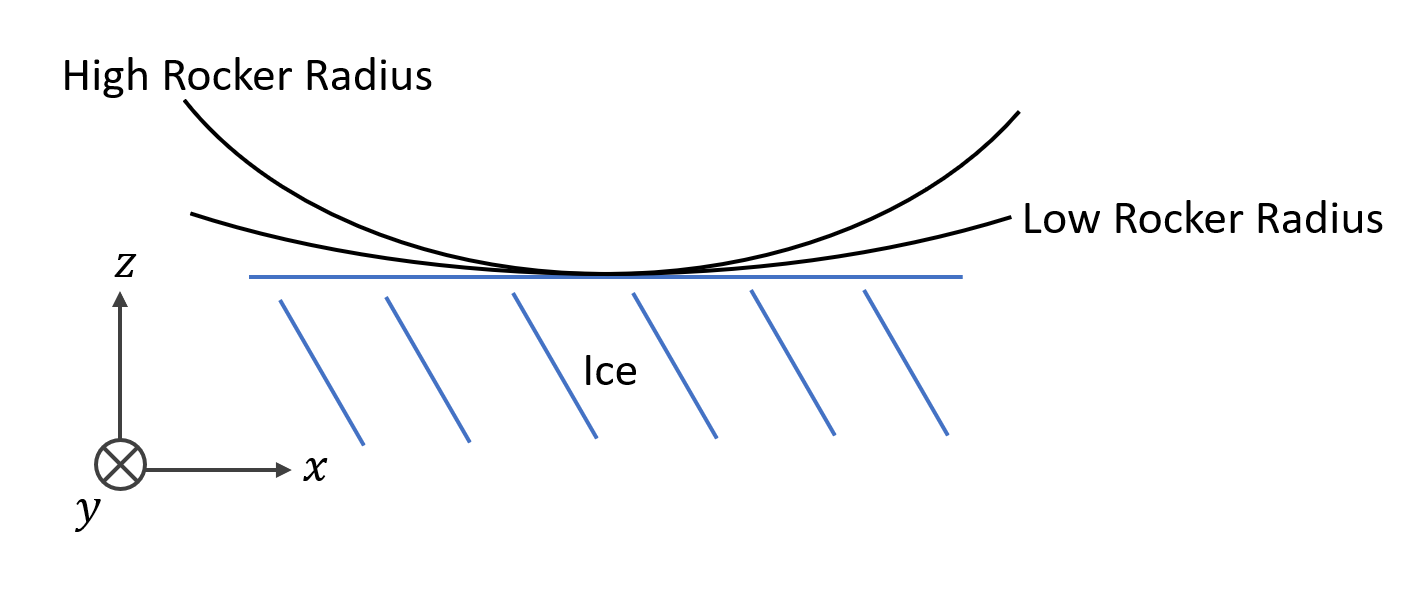}
        \caption{Longitudinal section}
    \end{subfigure}
    \hfill
    \caption{Visualization of the shape parameters of a luge steel which are changed for the experiment. The contact area is influences by the cross section and longitudinal section.}
    \label{fig:luge_steel}
\end{figure}

\begin{figure}
    \centering
    \begin{subfigure}[b]{0.49\textwidth}
        \centering
        \includegraphics[width=\textwidth]{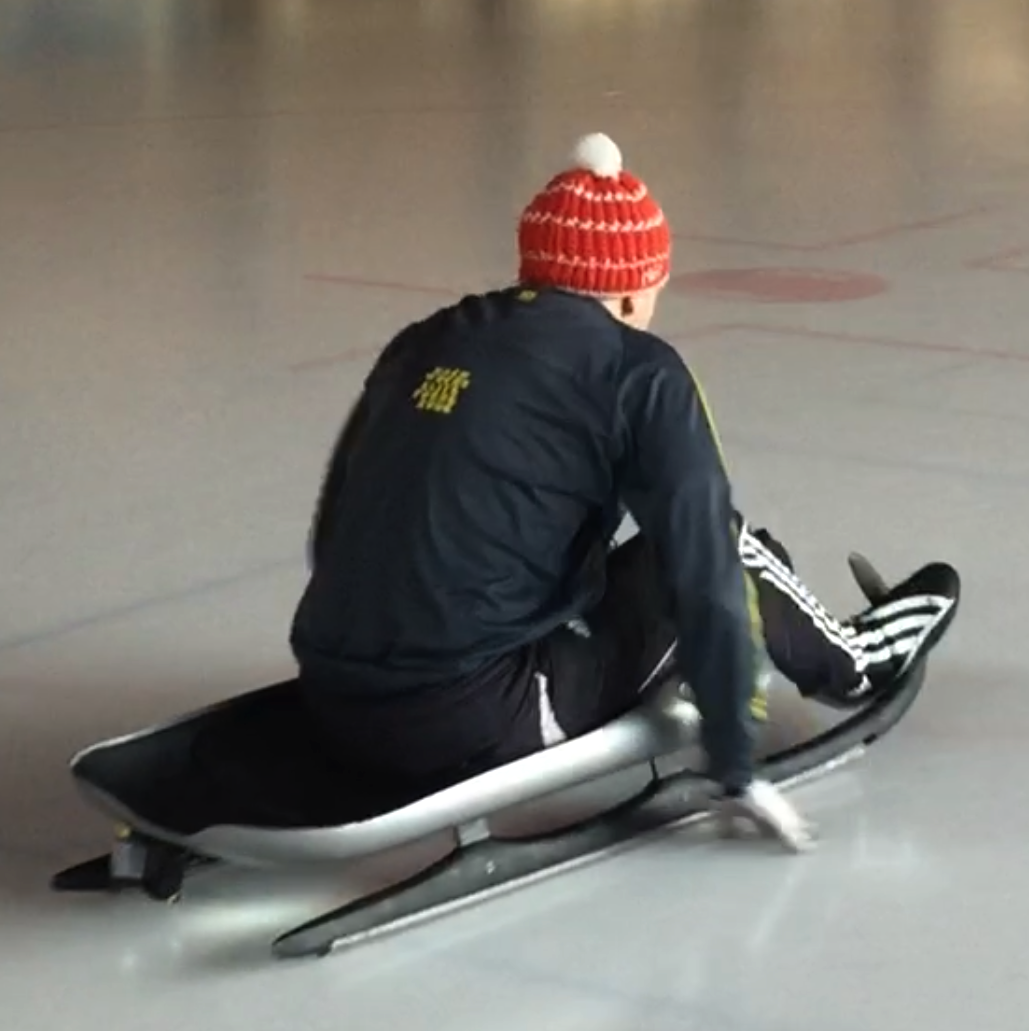}
        \caption{Acceleration phase}
    \end{subfigure}
    \hfill
    \begin{subfigure}[b]{0.49\textwidth}
        \centering
        \includegraphics[width=\textwidth]{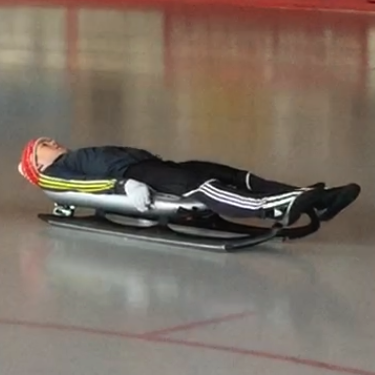}
        \caption{Gliding phase}
    \end{subfigure}
    \hfill
    \caption{Procedure of the ice house experiments. The athlete accelerates the sled similar to a luge starting phase. Friction is measured during the gliding phase.}
    \label{fig:glid_pic}
\end{figure}

Diverse samples are prepared for the investigation of ice friction, with special attention  given to the contact pressure. In contrast to most other publications, the pressure is altered by changing the contact area, instead of the normal force.  The samples are based on luge steels from the German national team with varying contact area geometries and are named Alpha, Beta, Gamma and Delta. Details are specified in Table \ref{tab:ice_house_res}. The numbers indicate different versions of a luge steel, which are generated by changing the set-up of luge steel and runner. The general shape of luge steels is sketched in \autoref{fig:luge_steel_shape}. \\
There are two variations regarding the pressure. The cross section is altered by tilting the luge steel on the runner, which is common practice in luge races. The rocker radius, i.e. the radius of the steels from a lateral point of view, is changed by bending the luge steels and compensating the occurring gap between luge steel and runner.  The specific geometries of the samples are confidential but not needed for implementing the methods described in the context of this paper. \autoref{fig:luge_steel}a shows an example for a sharp and round cross section of a luge steel and \autoref{fig:luge_steel}b a high and low curvature for the rocker radius, respectively  \\
The reached initial speed is $~8.5 \pm 0.5 km/h$. \autoref{fig:glid_pic}a shows that the sled is accelerated by strokes from the athlete, similar to a luge start.  Subsequently, the athlete lies down and the sled is in the gliding phase, during which the speed of the sled is continuously measured (\autoref{fig:glid_pic}b). 
%
To equalize the effect of a possible hill slope in the ice house,  all test runs are carried out in two opposite directions. Although, the ice house surface is designed to be flat, our sensitive experiment  yielded that there is a slope of approximately $0.12^\circ$.  The frictional force $F^\mathrm{ice}_\mathrm{x}$ is determined for an uphill and a downhill run and then averaged. The experimental conditions are shown in \autoref{ice_cond}.\\

\begin{table}[bt]
\caption{Ice house experiment conditions}
\label{ice_cond}
\begin{tabularx}{\textwidth}{XXXXXl}
\hline
Ice Temp.&Ambient Temp. &Humidity& Atmospheric Pressure\\
\hline
$-1.5^\circ C$&$2^\circ C$&$62\%$&$947hPa$\\
\hline
\end{tabularx}
\end{table}

\subsubsection{Data processing}
 In the experiment, we measure the speed of the sled during the gliding phase with the aforementioned measurement devices. The  force of ice friction can then be calculated  using energetic considerations which are detailed in this section. The total energy $E^\mathrm{tot}$ of a gliding sled can be described as the sum of  the potential energy $E^\mathrm{pot}$, the kinetic energy $E^\mathrm{kin}$ and the loss energy $E^\mathrm{loss}$ 

\begin{equation}
\label{eq:energy_tot}
\begin{split}
E^\mathrm{tot}=E^\mathrm{pot}+E^\mathrm{kin}+E^\mathrm{loss},\\
\text{with } E^\mathrm{loss}=E^\mathrm{aero}+E^\mathrm{ice}.
\end{split}
\end{equation}

The loss energy comprises of an aerodynamic part $E^\mathrm{aero}$ and an ice frictional part $E^\mathrm{ice}$. Assuming a distance $s$, an altitude  $h(s)$ and a velocity function $v(s)$, it results for a section from a point $s_0$ to $s_1$

\begin{equation}
\begin{gathered}
E^\mathrm{pot}=m \cdot g\cdot (h(s_1)-h(s_0)),\\
E^\mathrm{kin}=\frac{1}{2} \cdot m \cdot (v(s_1)^2-v(s_0)^2),\\
E^\mathrm{aero}=\int_{s_0}^{s_1} F^\mathrm{aero}_\mathrm{x} \: ds,
\end{gathered}
\end{equation}

with  $m$ being the mass, $g$ the earth's gravitational acceleration  and $F^\mathrm{aero}_\mathrm{x}$ the aerodynamic force.  It can be calculated with the drag equation by using $v$ and the ambient air conditions

\begin{equation}
\label{aero}
F^\mathrm{aero}_\mathrm{x}=\frac{C_\mathrm{x} A_\mathrm{x} \cdot v^2  \cdot \rho}{2}, \: \rho=\frac{p^\mathrm{air}}{R \cdot T},
\end{equation}

whereby $C_\mathrm{x}$ is the drag coefficient, $A_\mathrm{x}$ the cross section area, $p^\mathrm{air}$ the ambient pressure, $R$ the universal gas constant and $T$ the temperature of the air. The drag area for the combination of athlete and luge sled, i.e. $ C_\mathrm{x} A_\mathrm{x} $ was determined in the BMW windtunnel in Munich. 

\begin{figure}
\centering
\includegraphics[scale=0.16]{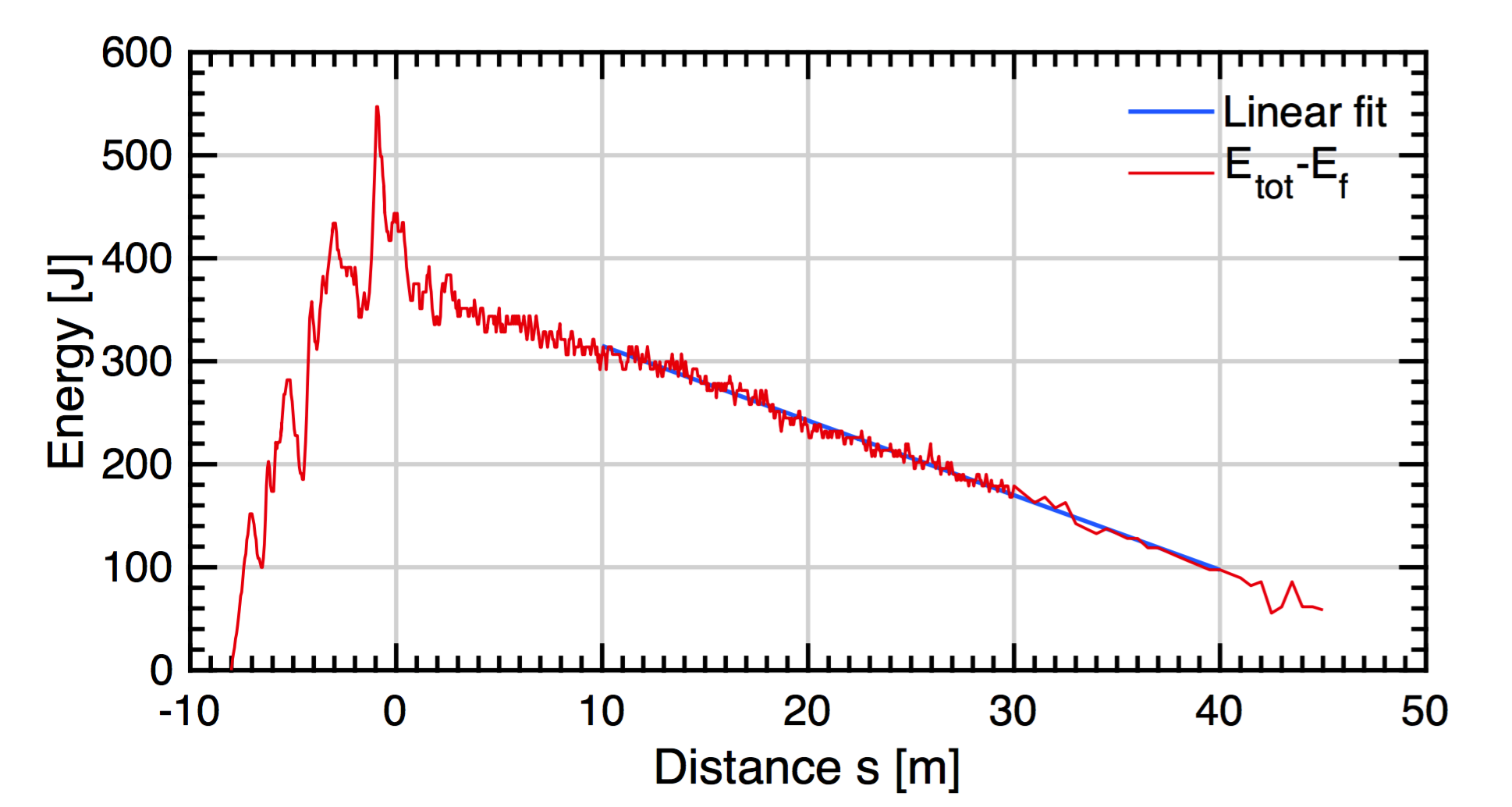} 
\caption[]{Energy diagram of a one run of the gliding experiment in the ice house. The negative slope represents the frictional force.} 
\label{fig:lin_fit}
\end{figure}

For the frictional force $F^\mathrm{ice}_\mathrm{x}$, it holds 
 \begin{equation}
 \label{eq:fric}
F^\mathrm{ice}_\mathrm{x}=\frac{d \: E^\mathrm{ice}}{ds}=-\frac{d \: (E^\mathrm{pot}+E^\mathrm{kin}+E^\mathrm{aero})}{ds}
 \end{equation}
by using the law of energy conservation.  The friction coefficient $\mu_\mathrm{x}$ for sliding friction  is defined as 
\begin{equation}
\label{mu_calc}
\mu_\mathrm{x}=\frac{|F^\mathrm{ice}_\mathrm{x}|}{|F^{}_\mathrm{z}|}=\frac{|F^\mathrm{ice}_\mathrm{x}|}{m \cdot g \cdot \cos(\kappa)},
\end{equation}
whereby  $F^{}_\mathrm{z}$ designates the normal force and $\kappa$ the hill slope angle. A linear least square fit was used to identify the slope of the energy function $\frac{dE}{ds}$ as shown in \autoref{fig:lin_fit}. To minimize unwanted effects from the laying-down or  sitting-up of the driver, only the middle section of the gliding phase was utilized.  

\subsubsection{Finite element simulation}
\begin{figure}
    \centering
    \begin{subfigure}[b]{0.9\textwidth}
        \centering
        \includegraphics[width=\textwidth]{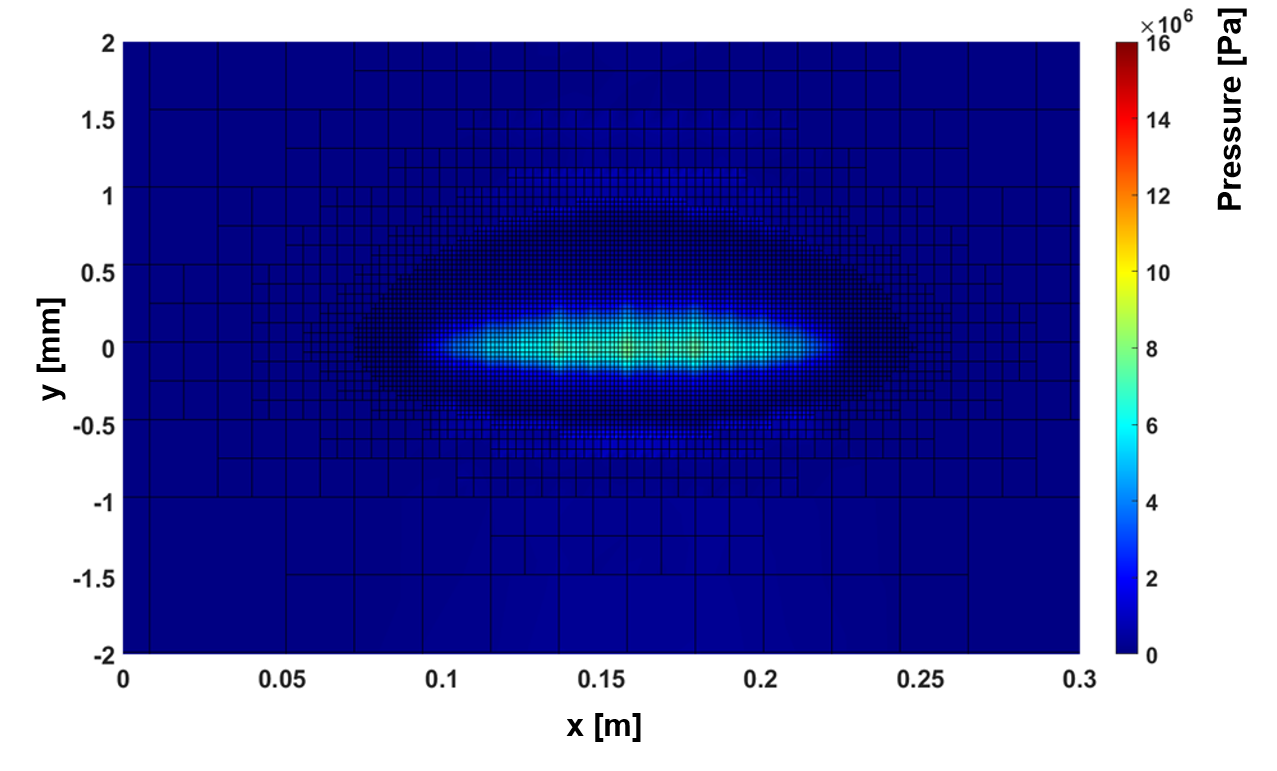}
        \caption{Alpha 1: round cross section}
    \end{subfigure}
    \hfill
    \begin{subfigure}[b]{0.9\textwidth}
        \centering
        \includegraphics[width=\textwidth]{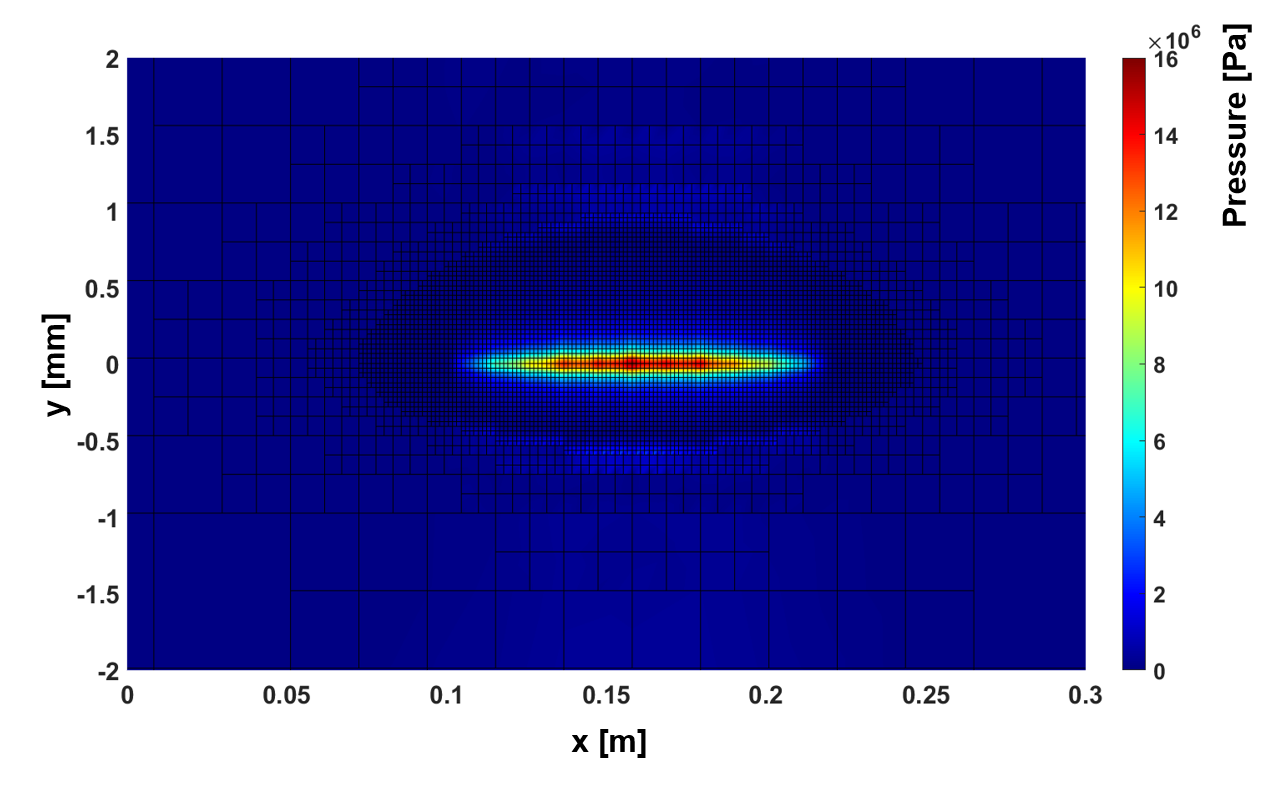}
        \caption{Beta 2: sharp cross section}
        
    \end{subfigure} 
    \hfill
    \caption[Ice surface pressure distributions]{Exemplary ice surface pressure distributions of two samples from the experiment. The sample Alpha 1 shows the lowest and Beta 2 the highest maximal pressure.}
    \label{fig:fem_iceh_res}
\end{figure}

A three-dimensional, linear elastic, static contact finite element (FEM) simulation is used to determine the contact pressure that a luge steel exerts on the ice. The simulation program is based on the MatLab Code  `Linear Elasticity' from \citet{alberty_matlab_2002}. In addition, the contact algorithm from \citet{schroder_projective_2008} is applied which is based on an accelerated projective SOR (Successive Over-Relaxation) approach. Hexahedral elements are introduced, boundary conditions are defined and the post-processing is altered to generate the desired output parameters. 
Luge specific measurement tools and algorithms are used to generate a three dimensional hull model of a luge steel. Other approaches would yield similar results for the geometry generation, therefore, we do not describe the procedure in detail. \autoref{fig:fem_iceh_res} shows exemplary results for two samples. The differences in contact pressure and width are clearly visible. \\
We note that the used FEM simulation does not aim at fully describing ice friction but should be interpreted as a pressure test. Thermal effects, dynamic effects or plowing the ice, e.g. described in \cite{Lozowski.2013,Lozowski.2014c}, are not considered. 

\subsubsection{Longitudinal friction equation}
The ice house experiment yields the friction coefficients of the samples and the FEM simulation the corresponding contact pressures. We fit a  quadratic least square fit to inter- and extrapolate the measured data. A quadratic approach seems to be a natural choice for us, because other works suggested that there exists a minimum  of $\mu_\mathrm{x}(p)$ \cite{hainzlmaier_tribologically_2005}.
For the longitudinal direction we propose the following model using the relation between $p$ and $\mu_\mathrm{x}$

\begin{equation}
\begin{gathered}
F_\mathrm{x} =  -\mu_\mathrm{x}   \cdot  F_\mathrm{z}  \cdot  \cos \big( \alpha \big),\\
\text{with } \mu_\mathrm{x}=10^{-3} \cdot \min \{\zeta_\mathrm{x} \cdot (B_\mathrm{x}\cdot p^2 -C_\mathrm{x} \cdot p +D_\mathrm{x}),~ E_\mathrm{x}\} .
\end{gathered}
\label{eq_Fx_alpha}
\end{equation}

Thereby, $B_\mathrm{x},C_\mathrm{x},D_\mathrm{x} $ are parameters from the quadratic fit. The parameter $E_\mathrm{x}$ is an upper threshold for $\mu_\mathrm{x} $, which is necessary since the quadratic fit itself would be unconstrained towards very high pressures. The cosine of the side slip angle $\alpha$ is included to have a reasonable transition from longitudinal to lateral friction in the range  of $-90^\circ \leq \alpha \leq 90^\circ$. For example, when the bob is sliding purely laterally ($\alpha = \pm 90^\circ$) we expect the longitudinal force to be zero. 
The parameter $\zeta_\mathrm{x}$ accounts for asperities. In the ice house and in the FEM simulation, the ice was  smooth without asperities. For these circumstances it holds $\zeta_\mathrm{x}=1$. However, on a real track there can be significant asperities, which we assume to lead to higher contact pressures due to a reduced contact area. This can be accounted for by setting $\zeta_\mathrm{x} > 1$. In fact, luge and bobsleigh athletes stated in personal communications that the runtimes on a bumpy track are significantly slower than on a smooth track for otherwise equal conditions. This supports our assumption of higher acting contact pressures, however, experiments should be conducted to obtain reliable data in the future.   \\

\subsection{Lateral friction model}
The accurate modeling of lateral friction is a key element for  realistic driving behavior in a driving simulator. The frictional forces are one order of magnitude higher than for the longitudinal direction, resulting in a lateral acceleration which can be measured with good accuracy.  Therefore, we can use a more direct approach than for the longitudinal friction and use track data for modeling. 

\subsubsection{Track experiments}

\begin{figure}
\centering
\includegraphics[width=0.80\textwidth]{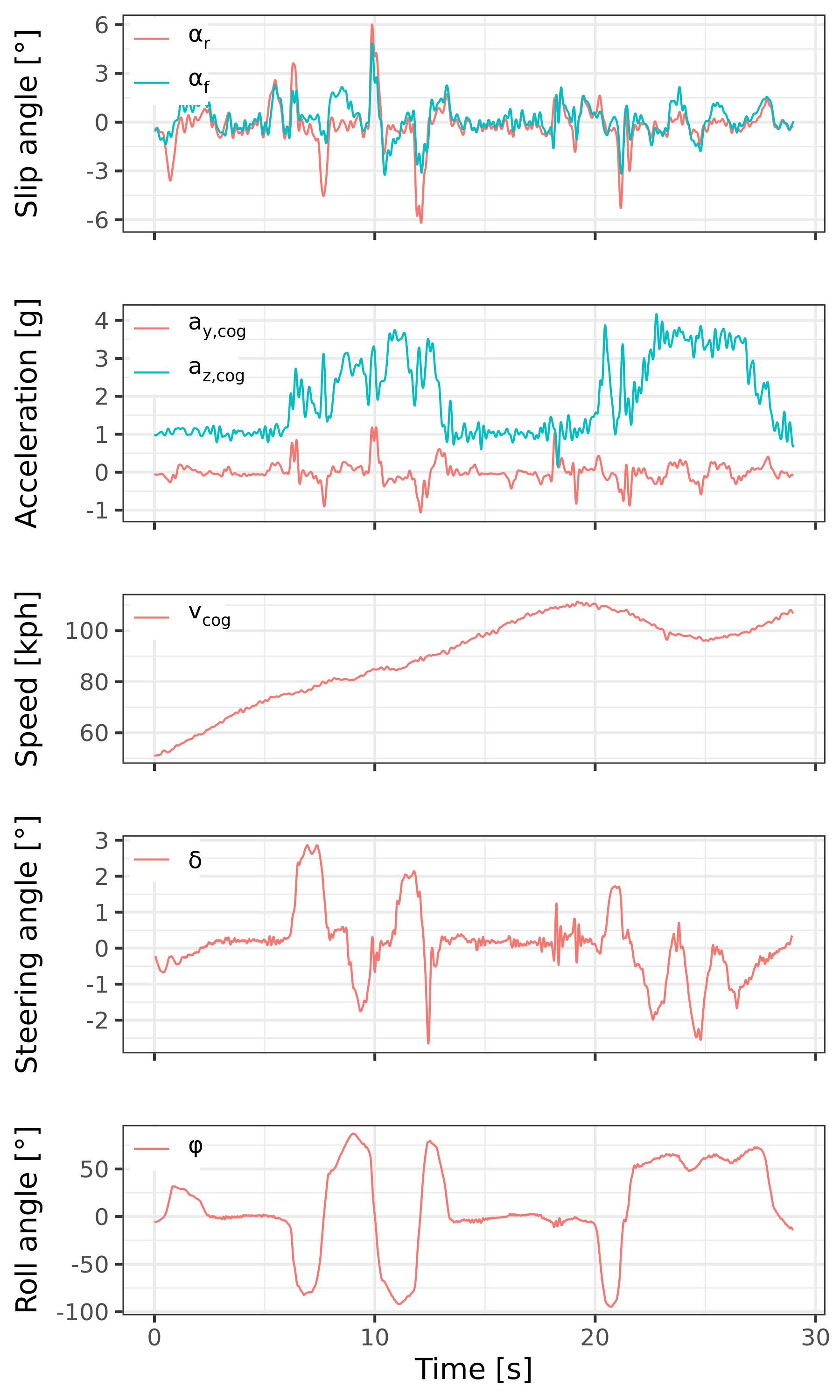}
\caption{Exemplary overview of selected measurement data in the time domain. The side slip angles for front and rear axle are on the top panel. On the second panel there are the accelerations followed by the speed, steering angle and the roll angle. The roll angle is obtained by integration of the roll rate and is well suited to distinguish the banked corners from each other. }
\label{fig_ts_overview}
\end{figure}

As summarized in \autoref{tab:track_data}, 44 runs with five different drivers were able to be obtained on the bob track in Königssee (Germany). \autoref{fig_ts_overview} shows an exemplary overview of the obtained signals in the time domain.  The run of an inexperienced driver is shown, which explains why the slip angles on the top panel are relatively high. For better visualization a 30 second extract of a run is shown. For the following analysis the complete run from start to finish is used. 
All signals are measured at the mounting position of the corresponding sensors. We transform the accelerations $\{a_\mathrm{x},a_\mathrm{y},a_\mathrm{z}\}$ to the center of gravity $\{a_\mathrm{x,cog},
a_\mathrm{y,cog},a_\mathrm{z,cog}\}$ using the rotational velocities, accelerations and distance of the sensor to the center of gravity $\{l_\mathrm{x},l_\mathrm{y},l_\mathrm{z} \}$ with the principles of rigid body dynamics

\begin{equation}
\begin{pmatrix}
a_\mathrm{x,cog}\\
a_\mathrm{y,cog}\\
a_\mathrm{z,cog}\\
\end{pmatrix} =
\begin{pmatrix}
a_\mathrm{x}\\
a_\mathrm{y}\\
a_\mathrm{z}\\
\end{pmatrix} -
\begin{pmatrix}
-\dot{\theta}^2-\dot{\psi}^2 & \dot{\varphi} \cdot \dot{\theta} - \ddot{\psi} & \dot{\varphi} \cdot \dot{\psi} + \ddot{\theta}\\
\dot{\varphi} \cdot \dot{\theta}  +\ddot{\psi}& -\dot{\varphi}^2-\dot{\psi}^2 & \dot{\theta} \cdot \dot{\psi}  -\ddot{\varphi}\\
\dot{\varphi} \cdot \dot{\psi}  -\ddot{\theta} &
\dot{\theta} \cdot \dot{\psi}  +\ddot{\varphi} & -\dot{\varphi}^2-\dot{\theta}^2
\end{pmatrix} 
\begin{pmatrix}
l_\mathrm{x}\\
l_\mathrm{y}\\
l_\mathrm{z}\\
\end{pmatrix}.
\label{eq_trans_accl}
\end{equation}
Thereby, we assume that the rotation speeds are constant for the whole body, i.e. rigid body dynamics can be applied. This is a simplification, especially for the front part of the bob because of the roll-split. \\
By geometric considerations, the side slip angle at other points along the x-axis can be calculated using the distance $l_\mathrm{s}$ from the sensor to the desired point and the yaw rate. For the side slip angle at the rear axle it holds
\begin{equation}
\alpha_\mathrm{r} = \alpha - \frac{\dot{\psi} \cdot l_\mathrm{s-r}}{v},
\end{equation}
with $\alpha$ being the side slip angle measured at the sensor's position. 
For the front axle, the steering angle $\delta$ has to be considered as well
\begin{equation}
\alpha_\mathrm{f} = \alpha - \frac{\dot{\psi} \cdot l_\mathrm{s-f}}{v} + \delta. 
\end{equation}

\begin{table}[h]
\caption{Data set overview for track experiments, the sample rate was downsampled to $100Hz$.}
\begin{tabularx}{\textwidth}{XXX}
\hline
Driver&Number of runs& Driver class\\
\hline
A1& 9 & Top level\\
A2& 12 & Top level\\
A3& 7 & Top level\\
B1& 12 & Advanced\\
C1& 4 & Junior\\
\hline
total & 44& \\
\hline
\end{tabularx}
\label{tab:track_data}
\end{table}

\subsubsection{One-track model of a bobsled}
\begin{figure}
 \centering
\begin{subfigure}[c]{0.81\textwidth}
\includegraphics[width=1\textwidth]{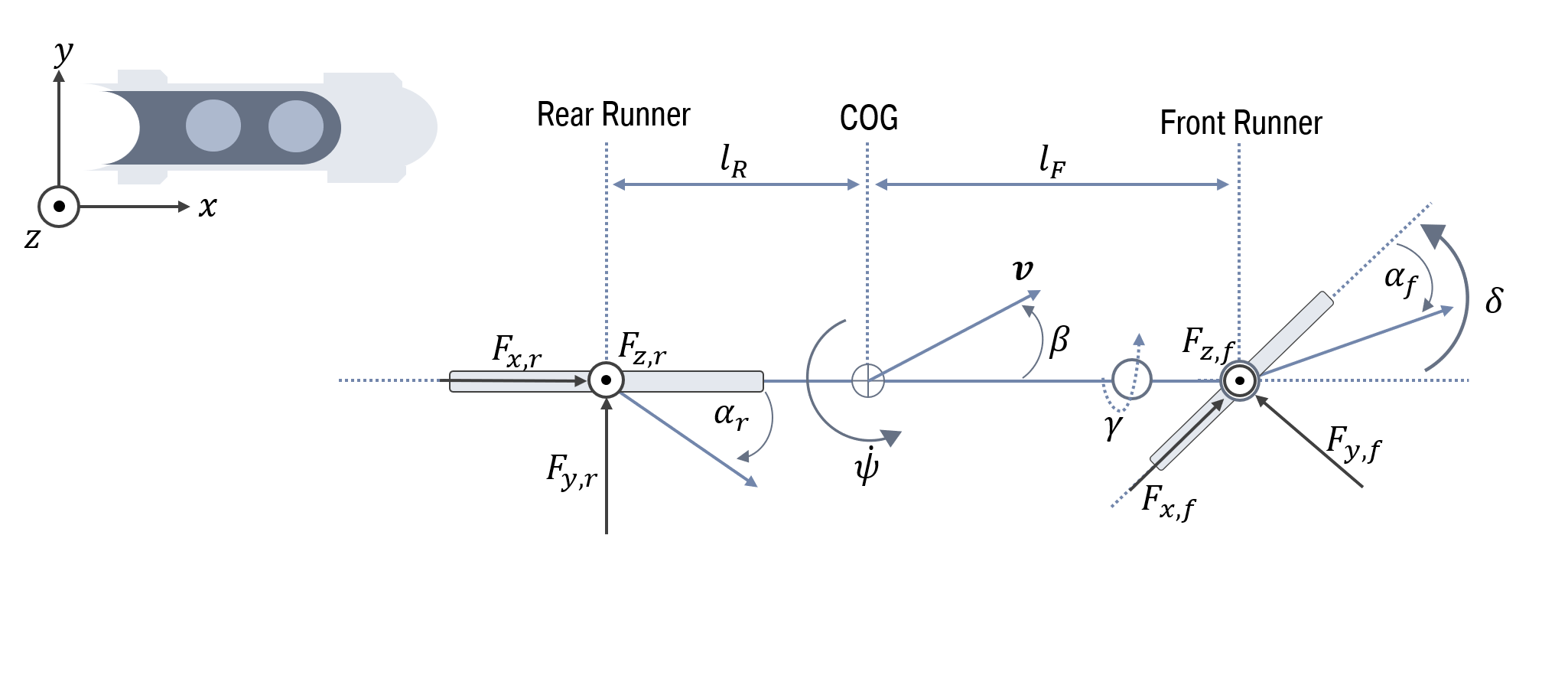}
\subcaption{Top view}
\end{subfigure}
\begin{subfigure}[c]{0.81\textwidth}
\includegraphics[width=1\textwidth]{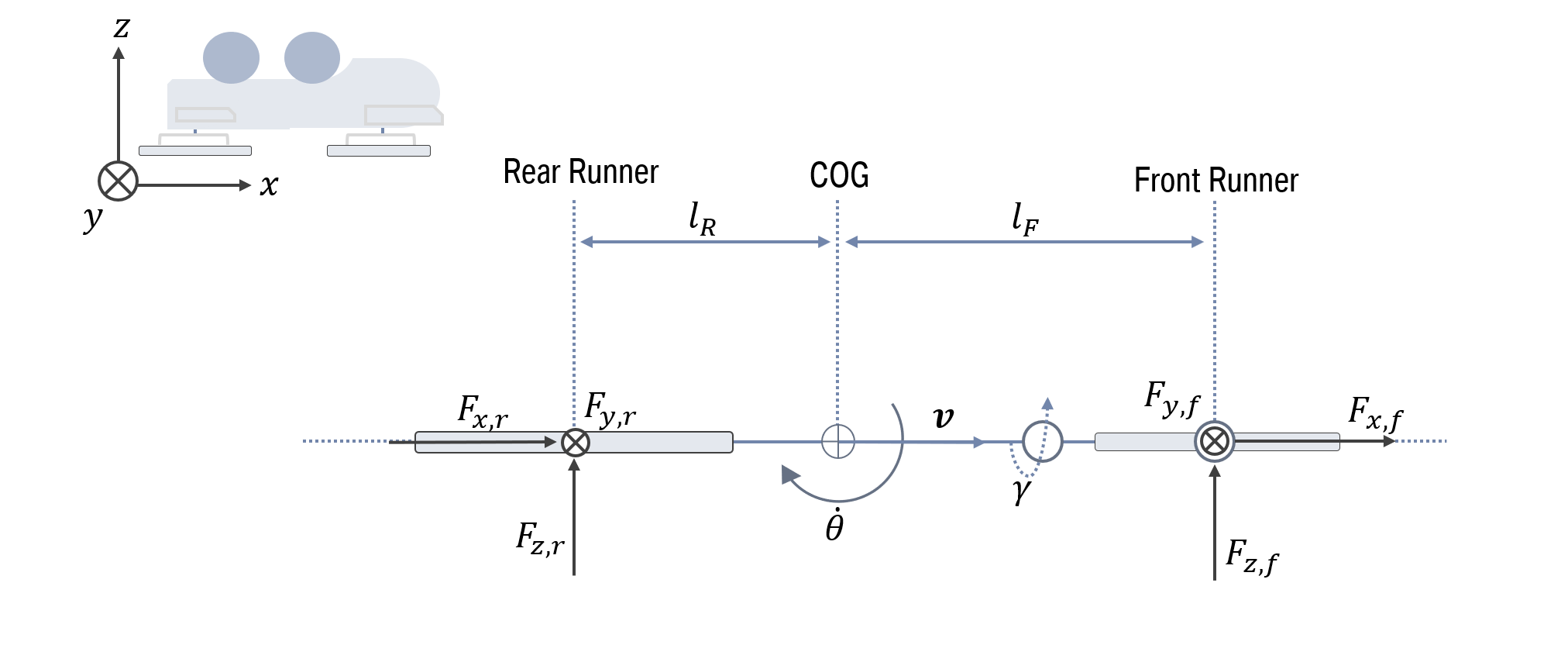}
\subcaption{Side view}
\end{subfigure}
\begin{subfigure}[c]{0.81\textwidth}
\includegraphics[width=1\textwidth]{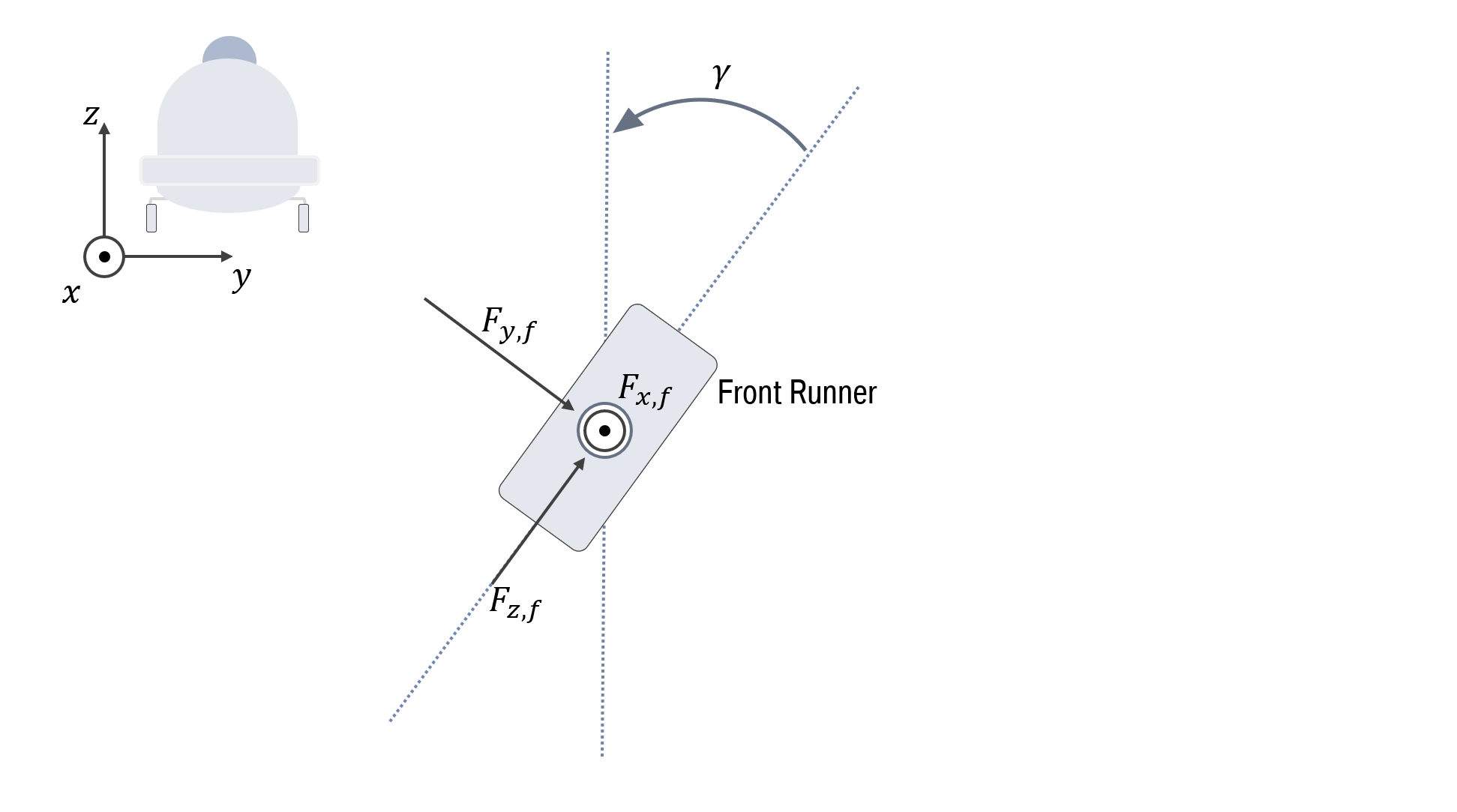}
\subcaption{Front view to show the tilt angle $\gamma$ of the front axle. }
\end{subfigure}
\caption{One-track model of a bobsled. The front axle rotates around its z-axis with the steering angle $\delta$, while the rear axle is static. Furthermore, there is a rotational joint to rotate the front around the x-axis by the roll-split angle $\gamma$ to represent the split of the front and rear part of a bobsled. Thereby, $\gamma$ includes both the rotation between the bobsled's front and rear part as well as the separate rotation of the front axle. }
\label{fig_one_track}
\end{figure}

For modeling lateral friction we need the forces at the runners which are not measured directly. Therefore, we create a  one-track model  of a bobsled. In an automotive one-track model, the left and right tire on each axle are combined to one tire \cite{Riekert.1940}. For the bobsleigh one-track model, we perform the same operation for the runners instead of the tires.  \autoref{fig_one_track} depicts the bobsleigh specific one-track model from three different angles. 

The model contains a single front and rear runner. The front runner can rotate around its z-axis ($\delta$), while the rear axle is static. In addition to a standard road car model, there is a rotary joint which rotates the front around the roll-split of the bob ($\gamma$). At the center of gravity  the bob has the velocity $v$ in the direction $\beta$. In consideration of the external forces $F_\mathrm{x,ext}$ and $F_\mathrm{y,ext}$ acting at the center of gravity (e.g. aerodynamic forces or wall contacts), the longitudinal, lateral and angular momentum equalities are

\begin{equation}
\begin{aligned}
m \cdot a_\mathrm{x,cog}~=~&F_\mathrm{x,r} +F_\mathrm{x,f0} + F_\mathrm{x,ext},
\end{aligned}
\label{eq:ImpulseLong}
\end{equation}

\begin{equation}
\begin{aligned}
m \cdot a_\mathrm{y,cog}~=~&F_\mathrm{y,r} +F_\mathrm{y,f0} + F_\mathrm{y,ext},
\end{aligned}
\label{eq:ImpulseLat}
\end{equation}

\begin{equation}
\begin{aligned}
m \cdot a_\mathrm{z,cog}~=~&F_\mathrm{z,r} +F_\mathrm{z,f0} ,
\end{aligned}
\label{eq:ImpulseVert}
\end{equation}

\begin{equation}
\begin{aligned}
J_\mathrm{zz} \cdot \ddot{\psi}~=~&l_\mathrm{F} \cdot F_\mathrm{y,f0} - l_\mathrm{R} \cdot F_\mathrm{y,r} +\\ 
&M_\mathrm{z,f}+M_\mathrm{z,r},
\end{aligned}
\label{eq:ImpulseYaw}
\end{equation}

\begin{equation}
\begin{aligned}
J_\mathrm{yy} \cdot \ddot{\theta}~=~&l_\mathrm{F} \cdot F_\mathrm{z,f0} - l_\mathrm{R} \cdot F_\mathrm{z,r} +\\ 
&M_\mathrm{y,f}+M_\mathrm{y,r}.
\end{aligned}
\label{eq:ImpulsePitch}
\end{equation}
As an approximation we assume that the torques $M_{y,f},~ M_{y,r},~ M_{z,f},~M_{z,r}$ are zero. The index `f0' denotes the point at the  front axle in the unrotated COG coordinate system, i.e. without the rotations $\gamma$ and $\delta$. It can be transformed to the front runner coordinate system `f' by first applying the rotation matrix 

\begin{equation}
\bm{A}_\gamma=
\begin{pmatrix}
1&0&0\\
0&\cos(\gamma) & -\sin(\gamma) \\
0& \sin(\gamma) & \cos(\gamma) \\
\end{pmatrix}
\label{eq_Ax}
\end{equation}
and afterwards the rotation $\delta$. Since $\delta$ is measured in the already rotated coordinate system (the front axle's steering axis rotates with the roll-split), the rotation axis of $\delta$ is dependent on $\gamma$ leading to the rotation matrix

\begin{equation}
\begin{gathered}
\bm{A}_{\delta}=
\begin{pmatrix}
\cos(\delta)&-\cos(\gamma) \sin(\delta)&-\sin(\gamma) \sin(\delta)\\
\cos(\gamma) \sin(\delta)& \sin^2(\gamma) \tilde{\delta} + \cos(\delta) & -\sin(\gamma) \cos(\gamma) \tilde{\delta} \\
\sin(\gamma) \sin(\delta)& -\sin(\gamma) \cos(\gamma) \tilde{\delta}&\cos^2(\gamma) \tilde{\delta} +\cos(\delta)\\
\end{pmatrix},\\
\text{with } \tilde{\delta} = 1-\cos(\delta).
\end{gathered}
\label{eq_Az}
\end{equation}

We define the transformation matrix from the `f0' to the `f' coordinate system $\bm{A}$ as

\begin{equation}
\bm{A} = \bm{A}_{\delta} \bm{A}_{\gamma}. 
\end{equation}

Consequently, it holds
\begin{equation}
\begin{pmatrix}
F_\mathrm{x,f}\\
F_\mathrm{y,f}\\
F_\mathrm{z,f}\\
\end{pmatrix} =
\bm{A} 
\begin{pmatrix}
F_\mathrm{x,f0}\\
F_\mathrm{y,f0}\\
F_\mathrm{z,f0}\\
\end{pmatrix}.
\label{eq_rotation_f0-f}
\end{equation}

and  for the inverse transformation
\begin{equation}
\begin{pmatrix}
F_\mathrm{x,f0}\\
F_\mathrm{y,f0}\\
F_\mathrm{z,f0}\\
\end{pmatrix} =
\bm{A}^{-1} 
\begin{pmatrix}
F_\mathrm{x,f}\\
F_\mathrm{y,f}\\
F_\mathrm{z,f}\\
\end{pmatrix} =
\bm{A}^{\text{T}} 
\begin{pmatrix}
F_\mathrm{x,f}\\
F_\mathrm{y,f}\\
F_\mathrm{z,f}\\
\end{pmatrix},
\label{eq_rotation_f-f0}
\end{equation}
with $\bm{A}^{\text{T}}=\bm{A}^{-1} $ since $\bm{A}$ is orthogonal.

\subsubsection{Limitations}
The measured data does not provide enough information for a two-track model, which would contain all four runners. Therefore, the effect of high roll accelerations and the resulting lateral load transfer are not considered. What is more, the acceleration transformation in \eqref{eq_trans_accl} does not account for the roll split of the bob. During high roll accelerations, the rear and front part of the bob  are accelerated to each other. Because of these properties, we exclude data for fitting of the friction model for which the roll acceleration is higher than a certain threshold. We choose a threshold of $100 ^\circ /s^2$ as a compromise between excluding the problematic high roll accelerations and keeping a sufficient amount of data from cornering phases.

\subsubsection{Lateral friction equation}
\label{sec_runner_model}
We fit a nonlinear regression model to the data set by adapting a part of the Pacejka `Magic Formula Tyre Model', which is an empiric model for tire road friction \cite{Pacejka.1992}.  Since the  bob glides in longitudinal direction, it is sufficient to model  pure lateral friction. We propose the following equation as a lateral friction model

\begin{equation}
\begin{gathered}
F_\mathrm{y} ~=~  \mu_\mathrm{y} \cdot \zeta_\mathrm{y} \cdot F_\mathrm{z}  \cdot  \sin \Bigg( C_\mathrm{y} \cdot  \arctan \bigg(B_\mathrm{y} \cdot \alpha - E_\mathrm{y} \cdot  \Big(B_\mathrm{y} \cdot  \alpha - \arctan \big( B_\mathrm{y} \cdot  \alpha \big) \Big) \bigg) \Bigg),\\
\text{with } B_\mathrm{y} ~= ~  \frac{K_\mathrm{y}}{C_\mathrm{y} \cdot \mu_\mathrm{y} \cdot \zeta_\mathrm{y} \cdot F_\mathrm{z}}.
\end{gathered}
\label{eq_Fy_alpha}
\end{equation}

Thereby, $\mu_\mathrm{y},~ \zeta_\mathrm{y},~ C_\mathrm{y},~K_\mathrm{y}$ and $E_\mathrm{y}$ are regression parameters, $\alpha$ and $F_\mathrm{z}$ predictor variables. The parameters are determined with the  \textit{MathWorks$^{\textrm{\textregistered}}$ MATLAB fitnlm} function which fits a specified nonlinear regression model to  given data.
We use a subset of the track data set (three days with similar bob settings and track conditions) for fitting the friction model. We predefine $E_\mathrm{y}$  to control the trend of $F_y$ towards higher absolute values of $\alpha$, since the measured range of slip angles was relatively small. We set $E_\mathrm{y}=0.99$ to slightly increase $F_y$ with increasing $\alpha$. This helps the driving simulation to be stable and predictable for higher slip angles. The parameter $\zeta_\mathrm{y}$ can be used as a tuning parameter, e.g. to account for asperities similar to the equation for the longitudinal friction. 

From the accelerations at the center of gravity, the forces at the front and rear runner need to be calculated to fit a friction model. A problem is that it is not possible to determine the split of $F_\mathrm{x,cog}$ into $F_\mathrm{x,f0}$ and $F_\mathrm{x,r}$ from the measured data with the one-track model. Furthermore, the force $F_\mathrm{x,cog}$ is very small compared to the other two forces and  sensitive to the alignment of the sensor which leads to a high relative measurement error. $F_\mathrm{x,r}$ is not needed to determine a lateral friction model, however, $F_\mathrm{x,f0}$ has to be known because of the required rotations $\gamma$ and $\delta$ to yield the forces at the front runner. To solve this problem we pre-define $F_\mathrm{x,f}$ in the front runner coordinate system using the already stated equation for longitudinal friction \eqref{eq_Fx_alpha}. In our case, this shows the need to consider both longitudinal and lateral friction in one model.   
According to  \eqref{eq_rotation_f0-f} it holds for the x-component of the matrix equation
\begin{equation}
{A}^{}_{(1,1)} F_\mathrm{x,f0} + {A}^{}_{(1,2)} F_\mathrm{y,f0} + {A}^{}_{(1,3)} F_\mathrm{z,f0} = F_\mathrm{x,f}.
\end{equation}
Consequently, $F_\mathrm{x,f0}$ is given by
\begin{equation}
 F_\mathrm{x,f0} = {A}^{-1}_{(1,1)} \big( F_\mathrm{x,f} - {A}^{}_{(1,2)} F_\mathrm{y,f0} - {A}^{}_{(1,3)} F_\mathrm{z,f0} \big). 
\end{equation}
With that, the necessary information is given to calculate all desired forces.

\subsection{Application in driving simulation}
\label{sec:driver_simulator}
\begin{figure}[]
\centering
\includegraphics[scale=0.55]{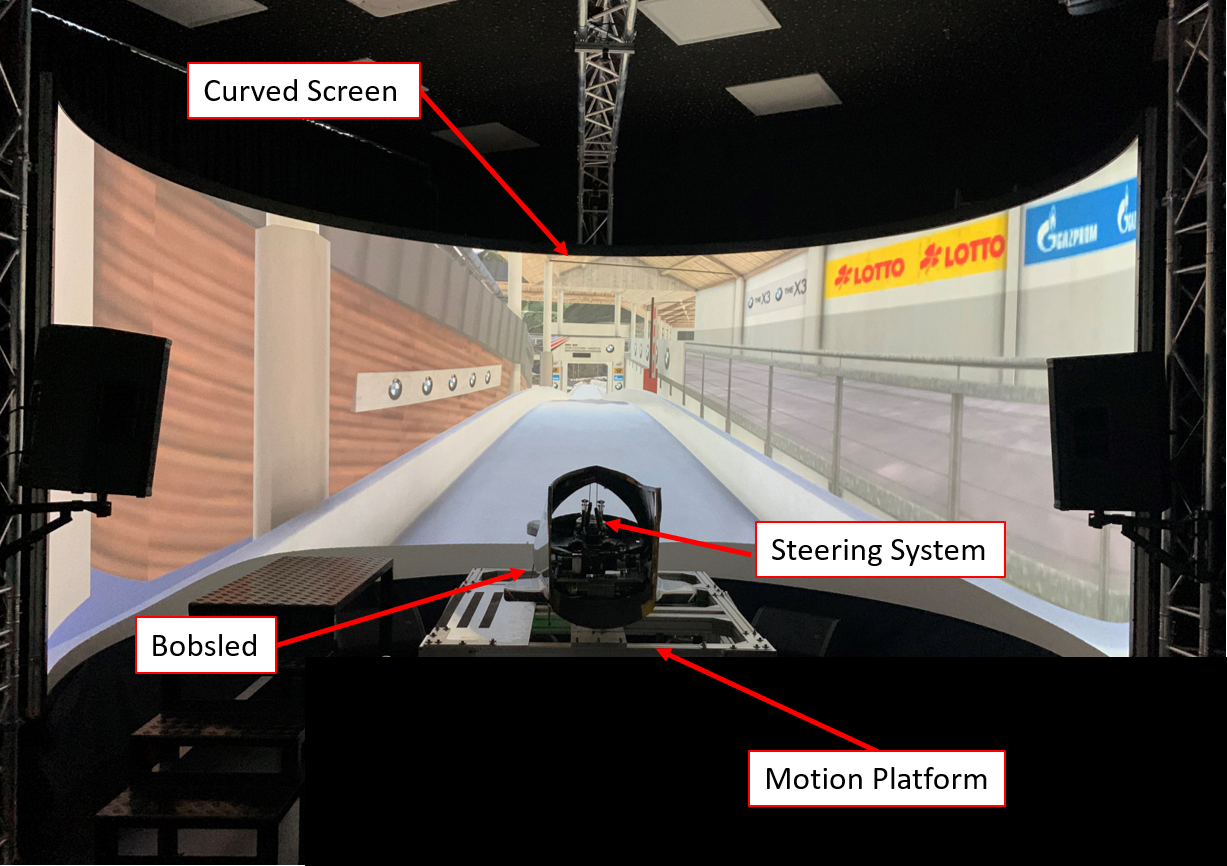} 
\caption{BMW bobsled driving simulator setup.}
\label{fig:dil}
\end{figure}

The  friction model here presented is deployed at the BMW bobsled simulator, which is serving as a  training ground for the German national bobsleigh team in preparation for the 2022 Olympic Winter games.  Its hardware is based on the BMW Motorsport simulator and is located in Munich \cite{Schwarzhuber.2020,Schleinitz.2021}. \autoref{fig:dil} shows the bob simulator setup. 

The mechanical assembly consists of a $210^\circ $ curved screen which surrounds a four degree-of-freedom motion platform on which an original bobsled is attached. The yaw rate $\dot{\psi}$, pitch rate $\dot{\theta}$ and roll rate  $\dot{\varphi}$ as well as heave motion (translation along the z-axis) can be rendered. The platform motion is controlled by a custom motion cueing algorithm. In addition, the steering mechanism of the bobsled is altered and includes a force feedback system.  \\
The  friction model is needed in the bob simulator to calculate the resulting forces at the runners based on  inputs from the simulation such as for example the normal forces and the slip angles at the runners. We use the equations \eqref{eq_Fx_alpha} and \eqref{eq_Fy_alpha} for the longitudinal and lateral frictional forces, respectively.\\
A precise modeling of the longitudinal friction force is required to achieve realistic run times and velocities on the track in the simulator. However, the pressure $p$ cannot be simulated by FEM as described in Section \ref{sec:fric_long} since the model needs to be realtime executable. Therefore, we propose a lookup for the pressure $p$ as a function of $F_\mathrm{z}$ and the track radius  around the y-axis of the bob $r_\mathrm{y,track}$, which can be calculated by using the pitch rate and velocity
\begin{equation}
r_\mathrm{y,track} = -\frac{v}{\dot{\theta}}.
\end{equation}
The lookup has to be generated for the front and rear runner, due to different geometries. As for luge steels, the cross section and rocker radius influence the pressure. We note that the geometries of bob runners provided by the German national team are confidential. Therefore  the results of the finite element simulations and the described lookup for the bob runners  are not published.  \\
The forces generated by lateral friction have a strong impact on vehicle dynamics  and are as a result decisive for a realistic driving behavior of the simulator. Therefore, the friction model is subject to continuous development based on data analysis and  driver feedback.  The data driven approach of this work aims to be a  good starting point for further development. In the first tests, the driving behavior was described to be realistic by professional athletes (World and Olympic champions in bobsleigh). However, the detailed analysis and validation of a driving simulator is an extensive task and will not be part of this publication. \\

\subsection{Application in Driver evaluation}
\label{sec:driver_method}
In bobsleigh, the achieved runtime on a track is decisive for winning races. Unsurprisingly, it has been the main criteria to evaluate bobsleigh drivers to this day. However, the runtime is influenced by many factors: 
\begin{itemize}
\item Start speed
\item Track conditions
\item Ambient conditions
\item Bobsled performance
\item Driver performance
\end{itemize}
Also, it is track specific and due to its strong sensitivity on outdoor conditions usually also session specific. Especially when comparing top drivers' performances, it is  necessary to have stable track and weather conditions and even then time comparisons are only viable when all runs are carried out within a short time frame.  Long-term and trend analysis to capture driver development are therefore hardly possible, if at all. \\
In contrast, we propose a method to evaluate  driver performance on the basis of the developed friction models for longitudinal \eqref{eq_Fx_alpha} and lateral friction \eqref{eq_Fy_alpha}  which can be applied to a data set containing multiple sessions, changing conditions and even different tracks. As stated above, the task of a bob driver is to achieve the fastest runtime. This can be accomplished by minimizing the ratio of traveled distance (from start to finish) and average velocity. 
Since the track in combination with the bobsled's driving dynamics restrict the possible racing line tightly, we focus on maximizing the average velocity in this work. This seems to be a sufficient approximation since no major deviations in the driven distance could be observed in the available data set. As stated in \eqref{eq:energy_tot} the only force accelerating a bobsled after the starting phase is the gravitational force. The driver can maximize the positive acceleration, and thereby the velocity, by minimizing the frictional losses which comprise of  ice friction and aerodynamic drag.\\

\subsubsection{Aerodynamics}
\label{sec:aero}
We want to model the aerodynamic drag force dependent on the bobsled's chassis side slip angle $\beta$.  An increase of the aerodynamic losses with increasing absolute  $\beta$ is expected due to an increased drag area $C_\mathrm{\tilde{x}}  A_\mathrm{\tilde{x}}$ in the driving direction   $\mathrm{\tilde{x}}$.  Thereby, we use  $\beta$ to introduce the coordinate system ($\tilde{x},\tilde{y},\tilde{z}$) where $\tilde{x}$ is aligned with the driving direction 
\begin{equation}
\label{eq_x_tilde}
\begin{pmatrix}
\mathrm{\tilde{x}}\\
\mathrm{\tilde{y}}\\
\mathrm{\tilde{z}}\\
\end{pmatrix} =
\begin{pmatrix}
\cos(\beta) & -\sin(\beta) & 0\\
\sin(\beta) & \cos(\beta) & 0\\
0 & 0 & 1\\
\end{pmatrix}
\begin{pmatrix}
\mathrm{x}\\
\mathrm{y}\\
\mathrm{z}\\
\end{pmatrix}. 
\end{equation}

 We do not have wind tunnel measurements for different side slip angles available, therefore we approximate the so-called yaw sensitivity using simulations from \citet{BelloMillan.2016} who provided a diagram of the drag area  $C_\mathrm{\tilde{x}}  A_\mathrm{\tilde{x}} $ over yaw angles for an `Ahmed body'. An Ahmed body depicts a simplified vehicle geometry which is usually used for simulation studies of automobiles. However, its shape is also comparable to a bobsled, it has a rounded front section and a straight lateral surface. The similarities should  be sufficient to yield a yaw sensitivity in the same order of magnitude. \\
From a diagram in the work of \citet{BelloMillan.2016} we extracted  an increase of the drag area $\Delta_{C_\mathrm{\tilde{x}} A_\mathrm{\tilde{x}}}$  over yaw angle in the range from $0-10^\circ$
\begin{equation}
\begin{gathered}
\Delta^\mathrm{Ahmed}_{C_\mathrm{\tilde{x}}  A_\mathrm{\tilde{x}}} = 3.2 ~\% / ^\circ,\\
\text{with } \frac{A_\mathrm{y}}{A_\mathrm{x}}   \approx 2.3.
\end{gathered}
\label{eq:ahmed_body}
\end{equation}

 
For a bobsled  we determined
\begin{equation}
\begin{gathered}
\frac{A_\mathrm{y}}{A_\mathrm{x}}   \approx 5
\end{gathered}
\end{equation}
which is $2.17$ times the  surface ratio in \eqref{eq:ahmed_body}. As a rough approximation we assume the relative increase to scale linearly with the relative surface area differences
\begin{equation}
\Delta^{}_{C_\mathrm{\tilde{x}}  A_\mathrm{\tilde{x}}} \approx 2.17 \cdot \Delta^\mathrm{Ahmed}_{C_\mathrm{\tilde{x}} A_\mathrm{\tilde{x}}} \approx 6.94  ~\% / ^\circ.
\end{equation}

This leads to a dependency of the drag area on $\beta$ 
\begin{equation}
C_\mathrm{\tilde{x}}  A_\mathrm{\tilde{x}} = C_\mathrm{x} A_\mathrm{x} + 0.0694  \cdot C_\mathrm{x} A_\mathrm{x}  \cdot \mid \beta \mid
\end{equation}

for a bobsled. The resulting aerodynamic force in driving direction $F^\mathrm{aero}_\mathrm{\tilde{x}}$ is then determined with \eqref{aero}. \\
It has to be noted that this is an approximation yielding only the order of magnitude of the effect. Wind tunnel experiments to determine the exact values should be conducted. What is more, the effect of track walls and the rotation rates of the bobsled are not considered in this approximation. Also the aerodynamic influence of the roll-split of the bob, which was examined by \citet{Sciacchitano.2018} is not considered.  To improve the aerodynamic model, computational fluid dynamics simulations should be conducted which include the stated parameters. 

\subsubsection{Driver evaluation criteria}
We assume that the optimal state, i.e.\ the state with minimal frictional losses, occurs when the bob is traveling in a straight line with no side slip angle at the runners or the center of gravity. For ice friction, this follows directly from our proposed friction models, since the lateral friction is much higher than the longitudinal friction. A similar effect is found for aerodynamics in Section \ref{sec:aero}. The consequence is that drifts should be avoided so as not to increase ice friction at both axles. Furthermore, the bobsled's chassis side slip angle $\beta$ should be zero to be aerodynamically optimal.\\
Apart from this qualitative statement, we can  calculate the effect of drifts for a quantitative analysis using the forces which act against the driving direction. 
Using the friction and  aerodynamic models, the force acting against the driving direction $F_\mathrm{\tilde{x}}$ can be determined using the coordinate transformation in \eqref{eq_x_tilde}.  It is compared to the force in x-direction in the body coordinate system $F_\mathrm{x}$, which we consider to be optimal. Of course, this optimum cannot be reached on a real track where steering and lateral forces are necessary, however, it is suited as a basis for comparison. We calculate $F_\mathrm{\tilde{x}}$ and $F_\mathrm{x}$  for  the front runner, rear runner and  for the aerodynamic drag using the coordinate transformation in \eqref{eq_x_tilde}. As an evaluation criteria for the driving style we integrate these forces, which leads to energies and compare the actual and minimal energy loss. First, we calculate the optimum, i.e. a theoretical, minimal loss energy  in x-direction

\begin{equation}
E^\mathrm{tot}_\mathrm{loss}=\int_{s_0}^{s_1} F_\mathrm{x,f}+F_\mathrm{x,r}+F^\mathrm{aero}_\mathrm{x} \,ds,
\end{equation}
whereby $s_0$ and  $s_1$ depict the start and finish at the track. Next, we propose several metrics based on relative loss energy increases

\begin{equation}
\begin{gathered}
\Delta E^\mathrm{ice}_\mathrm{loss,f}=\frac{\int_{s_0}^{s_1} F_\mathrm{\tilde{x},f0}-F_\mathrm{x,f} \: ds}{E^\mathrm{tot}_\mathrm{loss}},\\
\Delta E^\mathrm{ice}_\mathrm{loss,r}=\frac{\int_{s_0}^{s_1} F_\mathrm{\tilde{x},r}-F_\mathrm{x,r} \: ds}{E^\mathrm{tot}_\mathrm{loss}},\\
\Delta E^\mathrm{aero}_\mathrm{loss}=\frac{\int_{s_0}^{s_1} F^\mathrm{aero}_\mathrm{\tilde{x}}-F^\mathrm{aero}_\mathrm{x} \: ds}{E^\mathrm{tot}_\mathrm{loss}},\\
\Delta E^\mathrm{tot}_\mathrm{loss}= \Delta E^\mathrm{ice}_\mathrm{loss,f} + \Delta E^\mathrm{ice}_\mathrm{loss,r} +\Delta E^\mathrm{aero}_\mathrm{loss}. 
\end{gathered}
\end{equation}
We use the term $F_\mathrm{\tilde{x},f0}-F_\mathrm{x,f}$ for the front to accommodate for the losses which arise by  rotating the front runner with the angles $\gamma$ and $\delta$. The reason why we analyze the losses relative to the optimum is that the absolute energy difference is influenced by the velocity and normal forces and thus dependent on track conditions. \\
In summary, $\Delta E^\mathrm{tot}_\mathrm{loss}$ can be interpreted as an overall score for driver evaluation. The smaller the relative energy increase to the optimum, the better the driver. By analyzing the steering and side slip angles and comparing the losses arising at the front and rear axle we can distinguish and evaluate different driving styles. 
Note that  a lower energy loss does not necessarily  imply a faster runtime when comparing different runs. Energy lost at the beginning of a track has a greater influence on the runtime than the same amount of energy lost near the finish. If needed the runtime influence can be calculated from the energy losses and corresponding positions on the track. In the case of driver evaluation we assume the energy difference to be more meaningful than the runtime difference as it is applicable to different tracks and conditions. \\
Apart from drivers, this method can also be used to evaluate the characteristic of different bob tracks. For track specific bobsleigh adjustment it might be interesting to compare how much energy is lost on the front axle compared to the rear axle. 

\subsubsection{Limitations}
In principle this driving style evaluation should be independent of the conditions on the track. In contrast, the classic metrics such as top speed or runtime strongly depend on the conditions, e.g. \citet{Jansons.2021}  showed that weather conditions had a significant effect on the runtime even on a short starting track.  However, also for our proposed relative loss energy difference, there can be indirect influences from the conditions, e.g.  there may be circumstances (e.g. very low or high temperatures and a very hard or soft ice surface) for which it is harder to control the bob  leading to higher energy losses.  Though, for the ice conditions commonly present in training and competitions this effect should be small.     \\

\section{Results}

\subsection{Longitudinal friction model}
\begin{figure}[]
\centering
\includegraphics[scale=0.16]{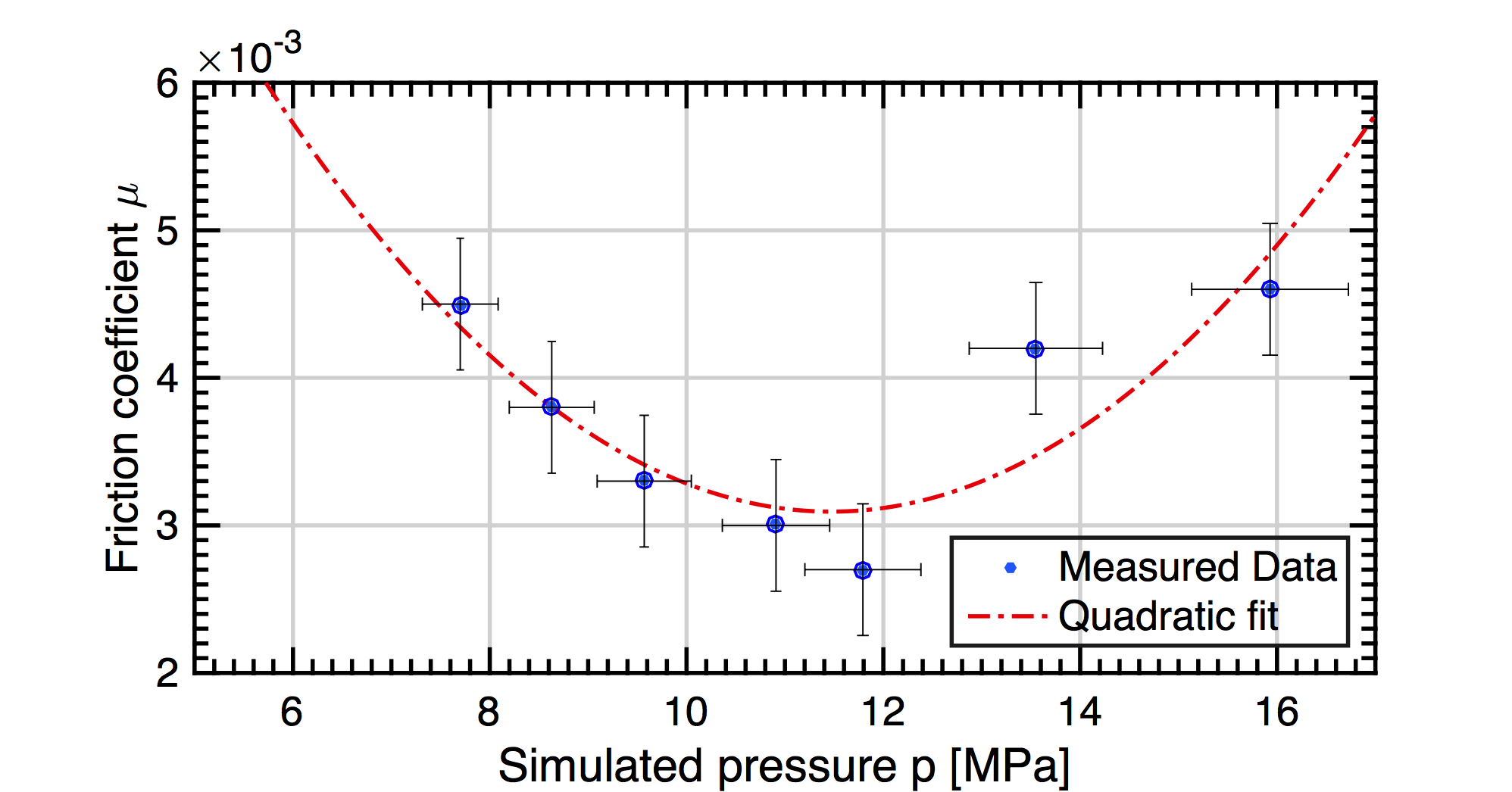} 
\caption{Pressure dependency of the longitudinal friction coefficient $\mu_\mathrm{x}$ for the ice house experiment specimens with luge steels.}
\label{fig:mu_pressure}
\end{figure}

The results of the ice house experiments  and the equivalent FEM simulations   are presented in Table \ref{tab:ice_house_res}. \autoref{fig:mu_pressure} shows a graphical overview of friction coefficient and pressure for the samples and the fit for $\mu_\mathrm{x}$ described in \eqref{eq_Fx_alpha}.
The friction coefficient continuously decreases from Alpha~1   to Gamma~2, where the minimum point is located. Friction rises from that point to Beta~1 and Beta~2. Since the Alpha samples exert little pressure, a decrease of the contact area by reducing the rocker radius from Alpha~1 to Alpha~2 leads to lower friction. Gamma~1 also belongs  to the low pressure regime as a pressure decrease to Gamma~3 raises friction and a pressure increase to  Gamma~2 reduces friction. In contrast, increasing the pressure from Beta~1 to Beta~2 leads to higher friction. Therefore, the Beta samples are allocated to the high pressure regime. The existence of an optimal pressure leading to a minimal friction coefficient is in accordance with the findings of \citet{hainzlmaier_tribologically_2005}. Mechanisms such as frictional heating and pressure melting could be responsible for a decrease in friction with increasing pressure \cite{hainzlmaier_tribologically_2005}. An explanation for an increase of friction for even higher pressures is that the mechanical stress on the ice surface becomes too high and the slider plows through the ice. The nonlinear pressure dependency of ice friction is also well known in the sport of luge. In summary, the resulting parameters for the longitudinal friction model in  \eqref{eq_Fx_alpha} are listed in \autoref{tab:runner_mode_x_params}.

\begin{table}[!htbp]
\caption{Experimental and simulation results for ice house experiments with luge steels.}
\begin{tabularx}{\textwidth}{XXXXX}
\hline
Specimen&Rocker Radius&Cross-Section& $\mu_\mathrm{x} [10^{-3}]$&$p_{max}$ $[MPa]$\\
\hline
Alpha 1&high&round, wide&4.5 $\pm 0.4$ &$7.7 \pm 0.4 $\\
Alpha 2&low&round, wide&3.8$\pm 0.4$ &$8.6 \pm 0.5$\\
\hline
Beta 1&high&sharp, narrow&4.2$\pm 0.4$ &$13.6 \pm 0.7 $\\
Beta 2&low&sharp, narrow&4.6 $\pm 0.4$&$16.0 \pm 0.9$\\
\hline
Gamma 1&high&normal, narrow&3.0$\pm 0.3$ &$10.9 \pm 0.6 $\\
Gamma 2&low&normal, narrow&2.7$\pm 0.3$ &$11.8 \pm 0.6 $\\
Gamma 3&high&normal, wide&3.3$\pm 0.3$ &$9.6  \pm 0.5$\\
\hline
\end{tabularx}
\label{tab:ice_house_res}
\end{table}

\begin{table}[h]
\caption{Parameters for the longitudinal friction equation \eqref{eq_Fx_alpha}}
\begin{tabularx}{\textwidth}{XXXXX}
\hline
$B_\mathrm{x}$ & $C_\mathrm{x}$& $D_\mathrm{x}$& $E_\mathrm{x}$& $\zeta_\mathrm{x}$\\
\hline
0.088 & 2.01 & 14.66 & 0.007 &1\\
\hline
\end{tabularx}
\label{tab:runner_mode_x_params}
\end{table}

As stated above, the results of the longitudinal friction coefficient for the bobsleigh runners, i.e. the simulated pressures of the bobsleigh runners cannot be published since the runner geometries are confidential. Therefore, we define $\mu_\mathrm{x}$ as a fixed value to make the following results of this paper easier to reproduce and not dependent on an undisclosed parameter 
\begin{equation}
\mu_\mathrm{x} \coloneqq 0.004.
\end{equation}
This value lies in the range of our simulated data and is in good agreement with the real life experiments from \citet{Poirier.2011} and  \citet{Irbe.2021}. 
  
\subsection{Lateral friction model}
\begin{figure}
\centering
\begin{subfigure}[c]{0.7\textwidth}
\includegraphics[width=0.9\textwidth]{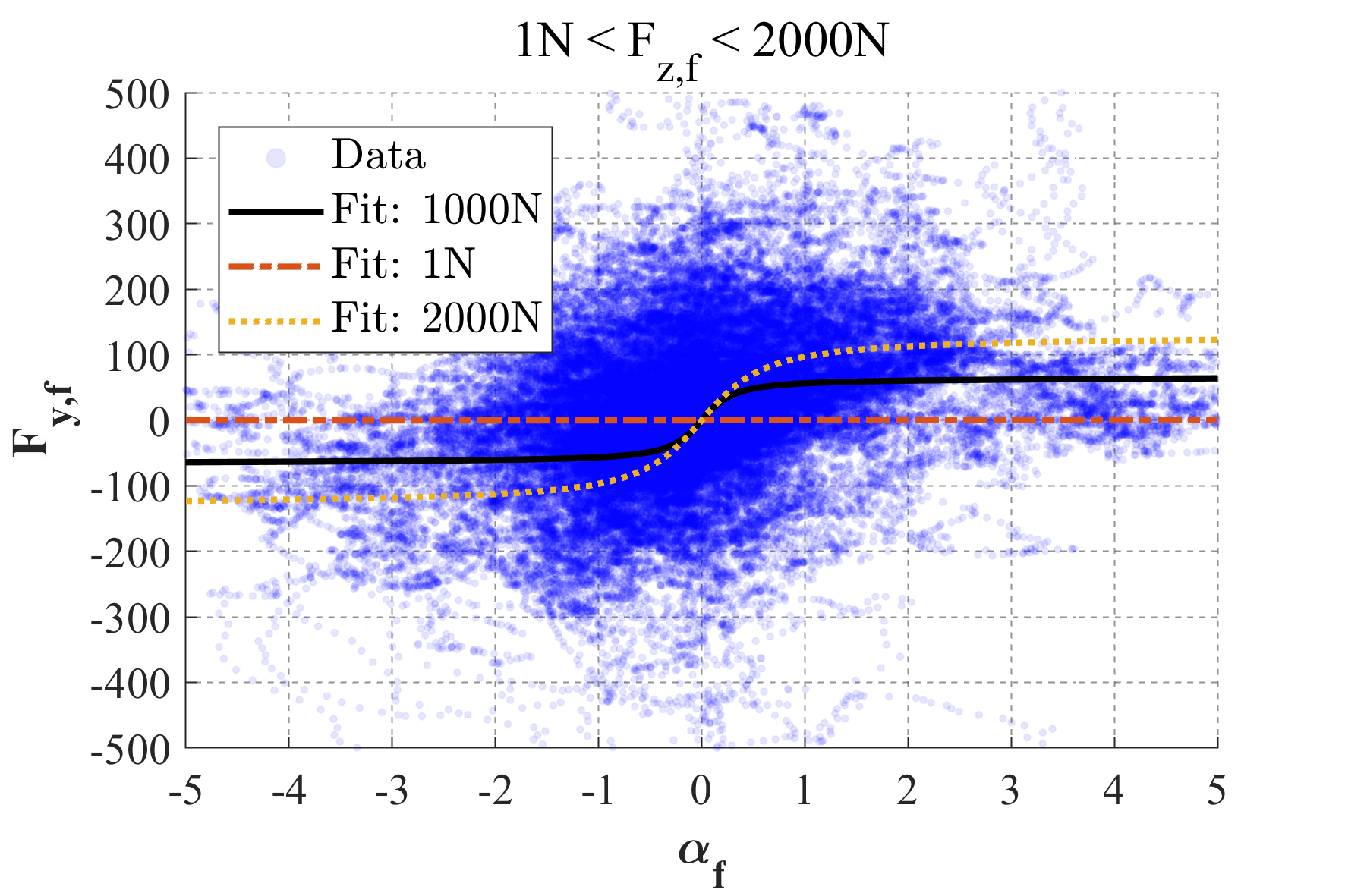}
\subcaption{}
\end{subfigure}
\begin{subfigure}[c]{0.7\textwidth}
\includegraphics[width=0.9\textwidth]{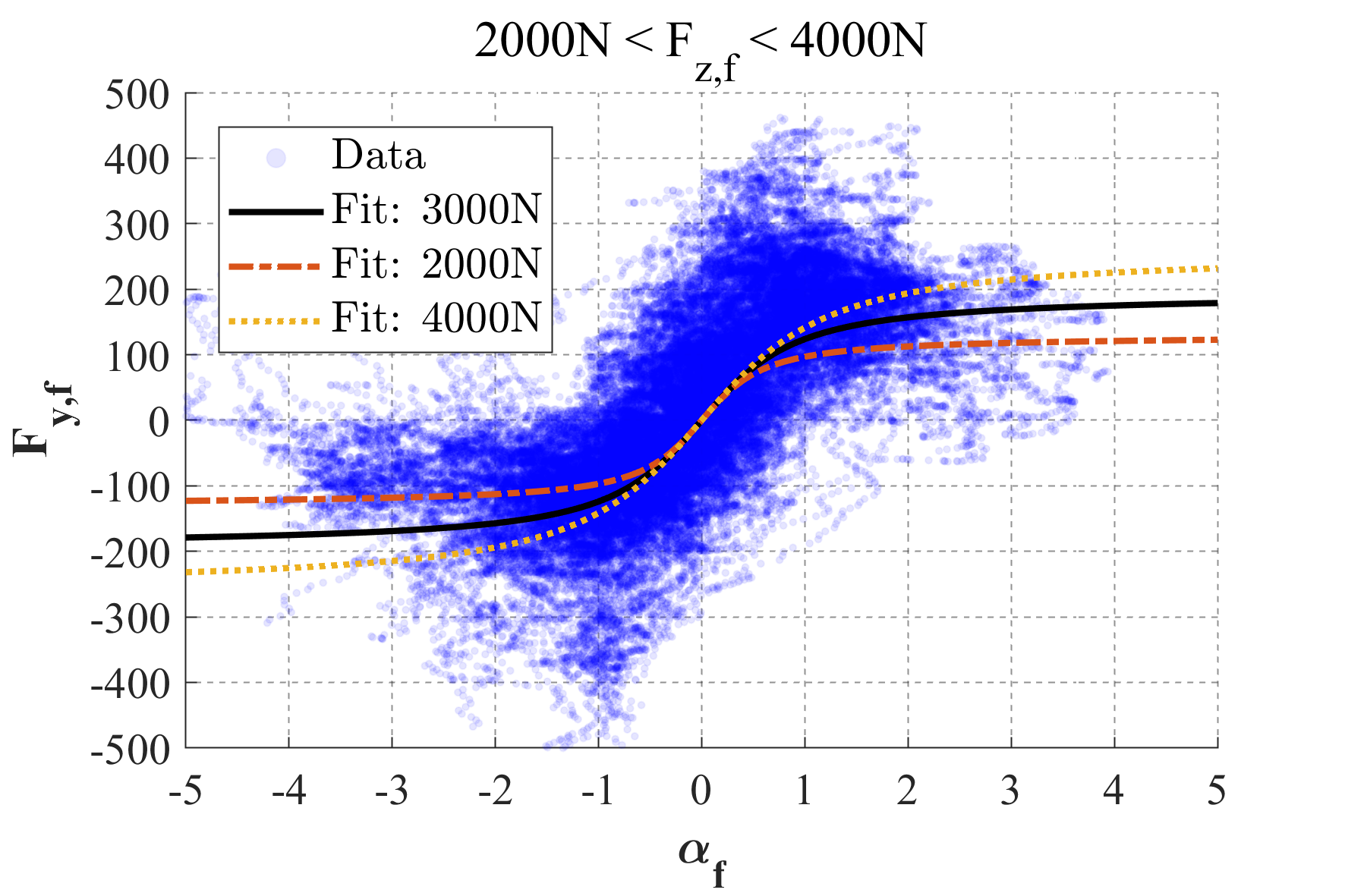}
\subcaption{}
\end{subfigure}
\begin{subfigure}[c]{0.7\textwidth}
\includegraphics[width=0.9\textwidth]{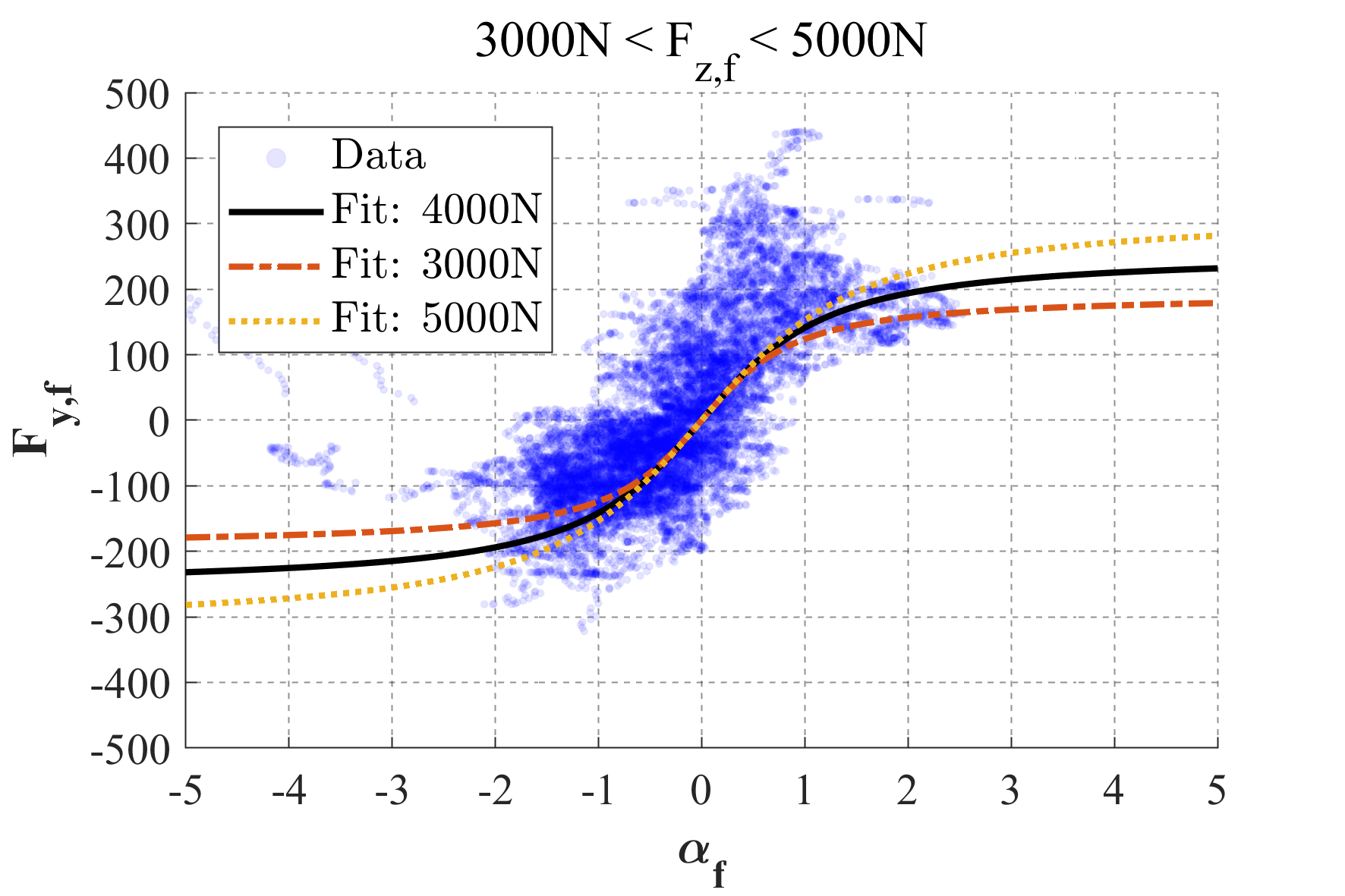}
\subcaption{ }
\end{subfigure}
\caption{Lateral friction model for the front runner  and measured data for different ranges of $F_\mathrm{z}$. }
\label{fig_fit_runner_front}
\end{figure}

\begin{figure}
\centering
\begin{subfigure}[c]{0.7\textwidth}
\includegraphics[width=0.9\textwidth]{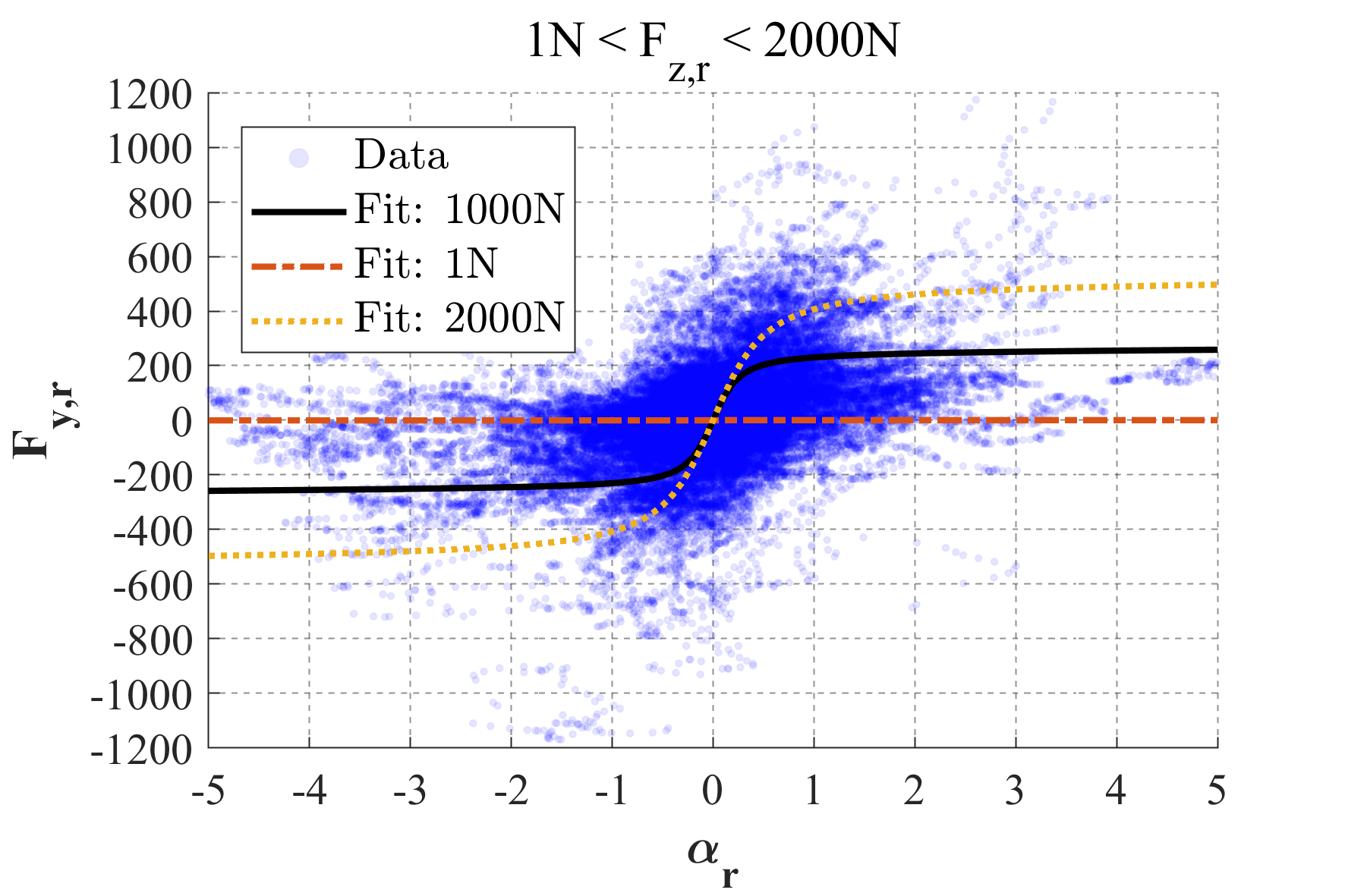}
\subcaption{}
\end{subfigure}
\begin{subfigure}[c]{0.7\textwidth}
\includegraphics[width=0.9\textwidth]{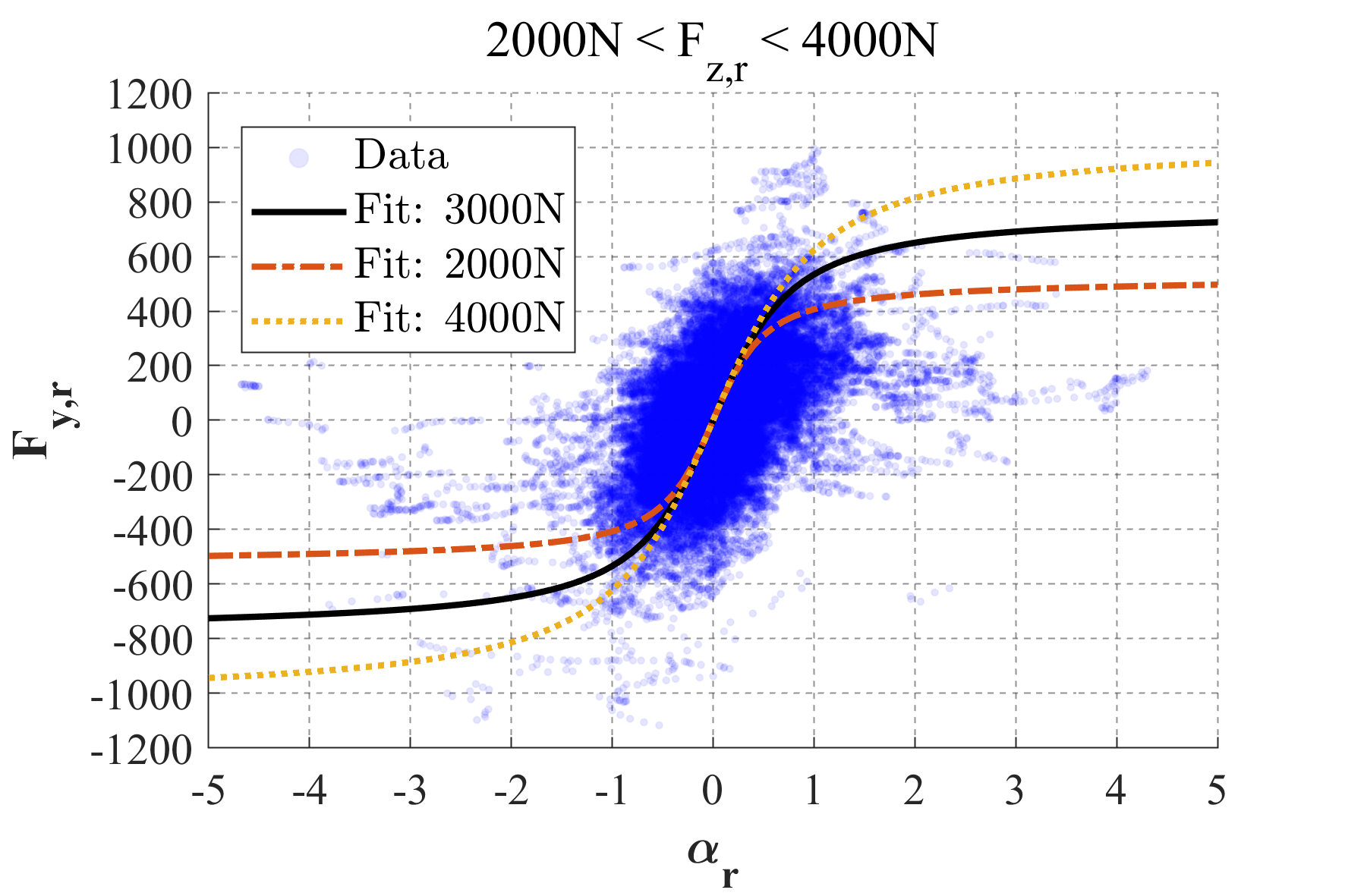}
\subcaption{}
\end{subfigure}
\begin{subfigure}[c]{0.7\textwidth}
\includegraphics[width=0.9\textwidth]{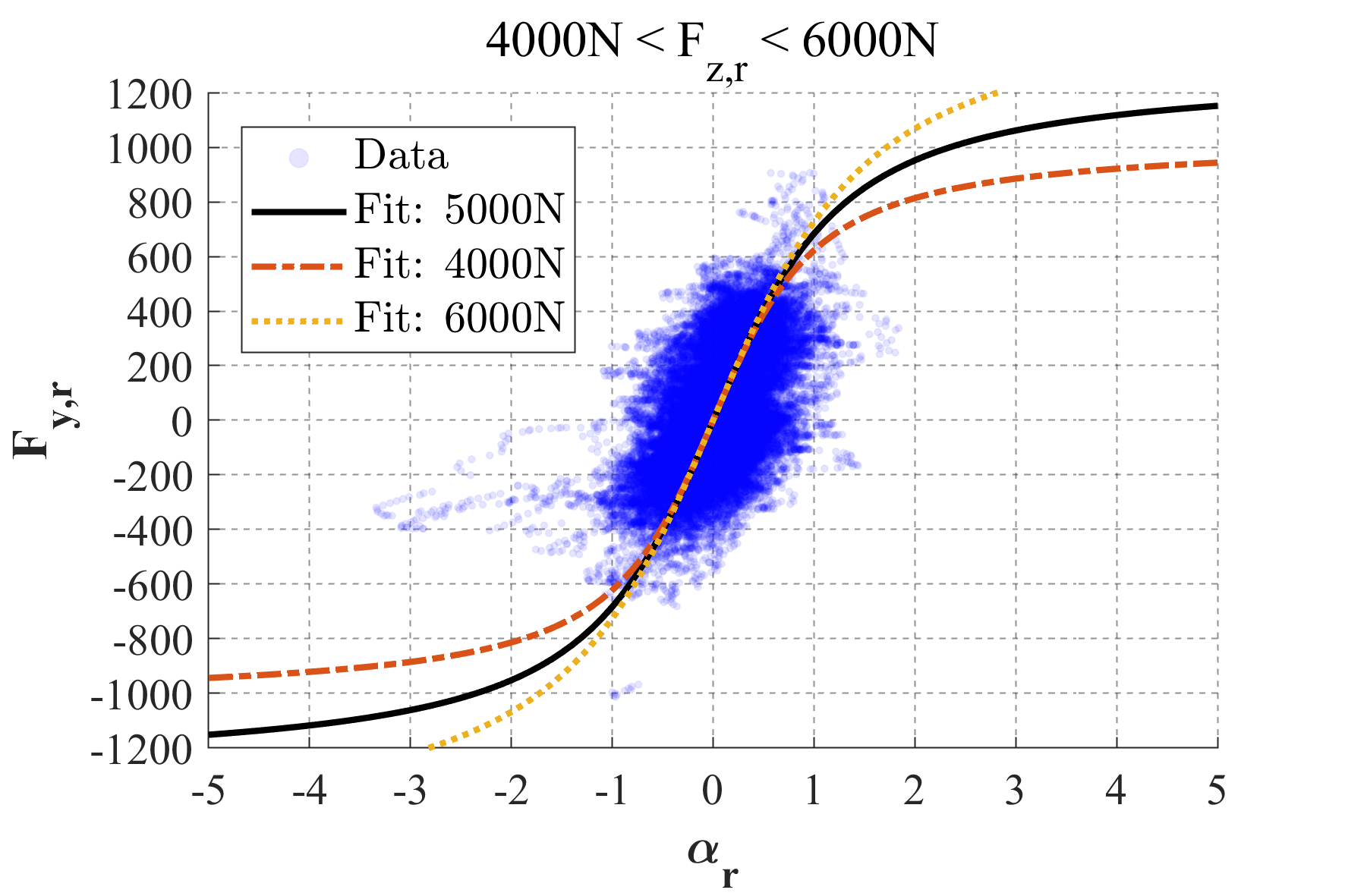}
\subcaption{}
\end{subfigure}
\caption{Lateral friction model for the rear runner  and measured data for different ranges of $F_\mathrm{z}$. }
\label{fig_fit_runner_rear}
\end{figure}

\autoref{fig_fit_runner_front} and \autoref{fig_fit_runner_rear} show the measured lateral force $F_\mathrm{y}$ over the slip angle $\alpha$ as well as the fitted models for different ranges of $F_\mathrm{z}$. The data (see \autoref{tab:track_data}) is filtered in the time  domain with a low-pass filter and a cutoff frequency of $20Hz$ to reduce noise. The spread of the data points is still high  and the correlation with the fits is relatively small for lower $F_\mathrm{z}$. For higher $F_\mathrm{z}$, there is a clearer correlation. In \autoref{fig_fit_runner_front}b, there  seems to be a maximum of $F_\mathrm{y,f}$ at $\alpha \approx 1^\circ$. A model with more regression parameters could fit this trend better, however, due to the high level of noise we  deliberately choose a model that underfits the data slightly.\\
The rear runner has a distinctly higher cornering stiffness, i.e. it produces more lateral force at a given slip angle than the front runner. The range of the measured data is even smaller, since bob drivers avoid too high slip angles at the rear axle due to dynamic stability (high drifts at the rear lead quickly to uncontrolled oversteer). 
In summary, the obtained parameters for the lateral friction models are listed in \autoref{tab:runner_mode_y_params}.

\begin{table}[h]
\caption{Lateral friction model parameters for front and rear runner. $E_\mathrm{y}$ was predefined to the stated value.}
\begin{tabularx}{\textwidth}{XXXXX}
\hline
Runner& $\mu_y \cdot \zeta_y$& $C_y$& $E_y$& $K_y$ \\
\hline
front& 2.577 & 0.024 &0.99& 10522\\
rear& 3.288 &0.076 &0.99 & 49776 \\
\hline
\end{tabularx}
\label{tab:runner_mode_y_params}
\end{table}

\subsubsection{Validation}
\begin{figure}
\centering
\includegraphics[width=0.9\textwidth]{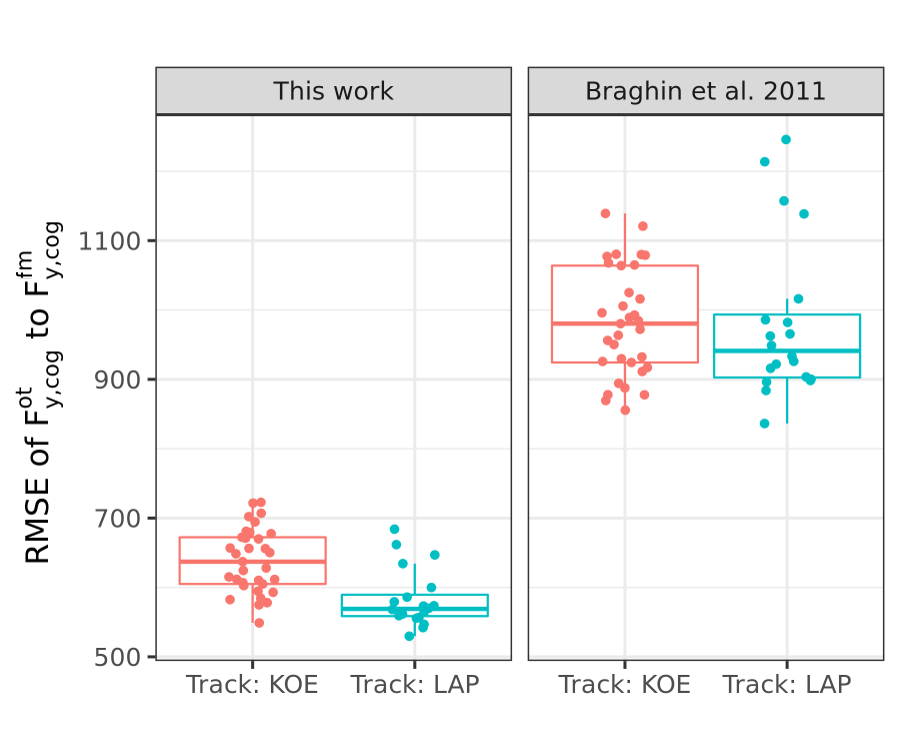}
\caption{RMSE comparison between the lateral force determined from measurement and the one track model $F^\mathrm{ot}_\mathrm{y,cog}$ and the calculated force from the runner model $F^\mathrm{fm}_\mathrm{y,cog}$ on the tracks in Königssee (KOE) and La Plagne (LAP). Each point accounts for a complete run. The model proposed by \citet{Braghin.2011} is shown for comparison.}
\label{fig_Fy_rmse}
\end{figure}

\begin{figure}
\includegraphics[width=0.95\textwidth]{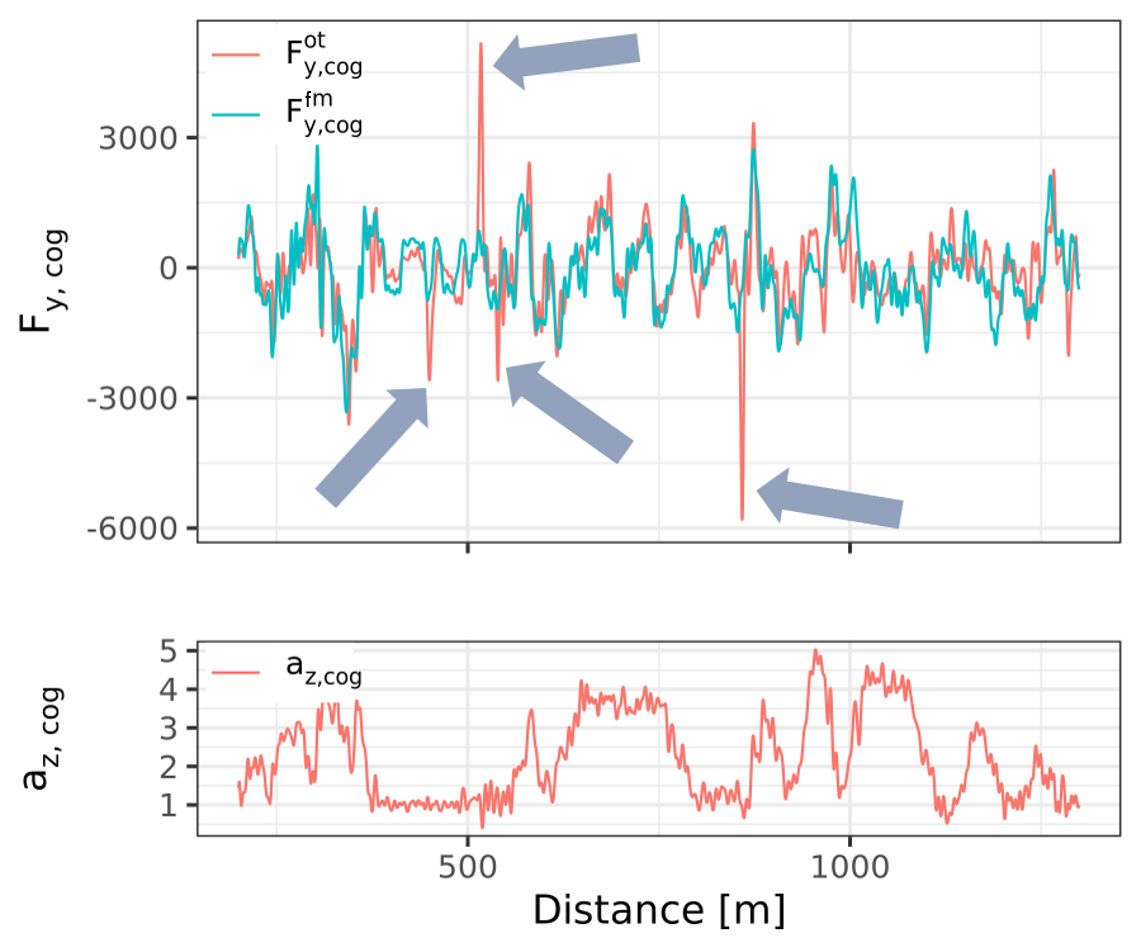}
\caption{$F^\mathrm{ot}_\mathrm{y,cog}$ and $F^\mathrm{fm}_\mathrm{y,cog}$ for the track in Königssee. The arrows indicate places where the bobsled hit a wall.}
\label{fig_Fy_koe}
\end{figure}

To validate the lateral friction model and to check how general the model is, we compare the lateral force at the center of gravity of the one-track model with the index `ot' $F^\mathrm{ot}_\mathrm{y,cog}$ with the force $F^\mathrm{fm}_\mathrm{y,cog}$ yielded by the friction model with the index `fm'. Using \eqref{eq_trans_accl}, the force of the one-track model comes directly from measurement. For the friction model, the whole calculation chain is necessary, i.e. determining $F_\mathrm{z,f}$, $F_\mathrm{z,r}$, $\alpha_\mathrm{f}$ and $\alpha_\mathrm{r}$, applying the friction model with the specified  parameters, transforming the determined forces at the front runner and finally summing up the forces at the front and rear runner. \\
\autoref{fig_Fy_rmse} shows the root mean square error (RMSE) between the signals. The  data used for validation was not used for fitting the model. We compare two different bob tracks, Königssee (KOE) and La Plagne (LAP), from which no data was used for model fitting. Interestingly, the RMSE in La Plagne is even lower than in Königssee. A contribution to the higher error of Königssee can be seen in the time domain in \autoref{fig_Fy_koe}. The blue arrows indicate areas where the bobsled hit the wall, which resulted in an external force which is not caused by runner-ice friction and therefore not covered by the friction model. In Königssee, hitting the wall is indeed the correct racing line, especially for the long straight around $500m$. In the so-called `Labyrinth' of the track, wall contact  is also common (arrow at around $800m$). In La Plagne, wall contact is limited to rare driving errors, which is an explanation for the lower RMSE. 
To sum up, the friction model seems to work well on unseen data and also on a completely different track. Apart from the model developed in this work, the RMSE of the proposed model from \citet{Braghin.2011} in \eqref{eq_fric_bra} is shown. The error is considerably higher on both tracks. We opted to compare the models with the provided parameters. Fitting their model to our data, especially if front and rear axle were to be be fitted separately, would certainly improve the accuracy.

\subsection{Application: Driver evaluation}

\begin{figure}
\centering
\begin{subfigure}[c]{0.8\textwidth}
\includegraphics[width=0.9\textwidth]{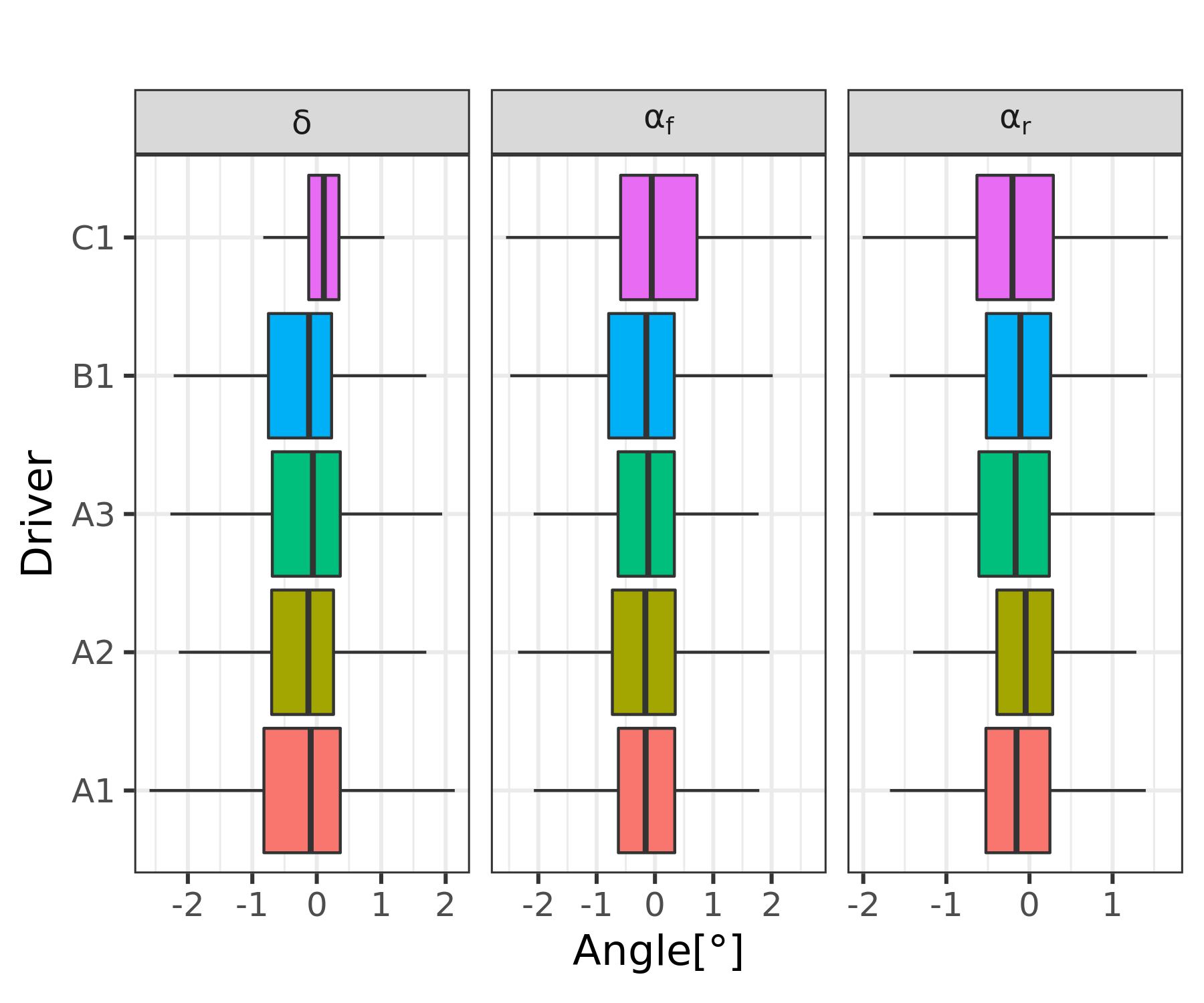}
\subcaption{Steering and side slip angle overview}
\end{subfigure}
\begin{subfigure}[c]{0.8\textwidth}
\includegraphics[width=0.9\textwidth]{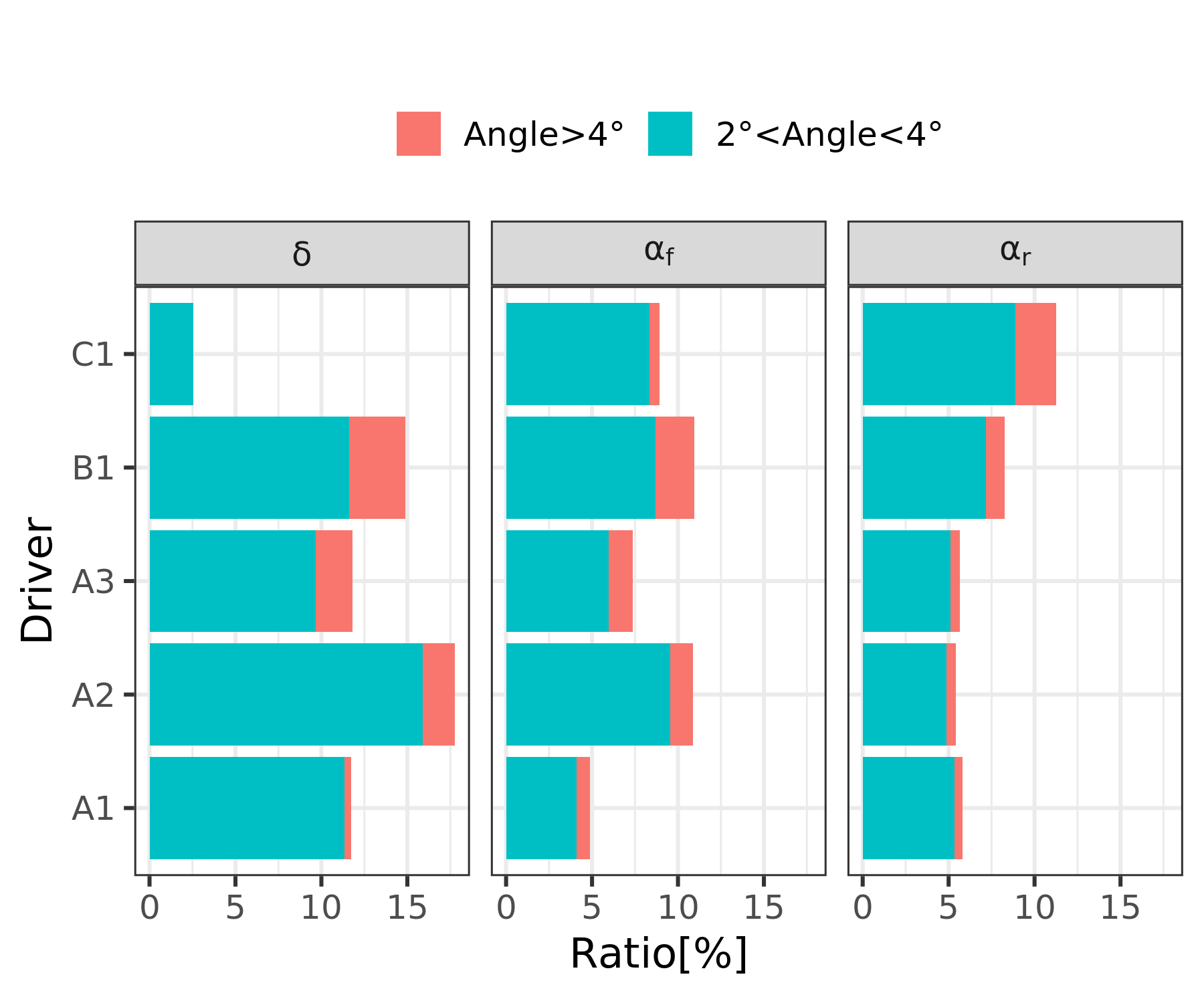}
\subcaption{Overview of high steering and side slip angles.}
\end{subfigure}
\caption{Driver comparison using measured signals. Clearly, Driver C1 steers much less than all other drivers. }
\label{fig_boxplot_overview}
\end{figure}

\begin{figure}
\centering
\includegraphics[width=0.88\textwidth]{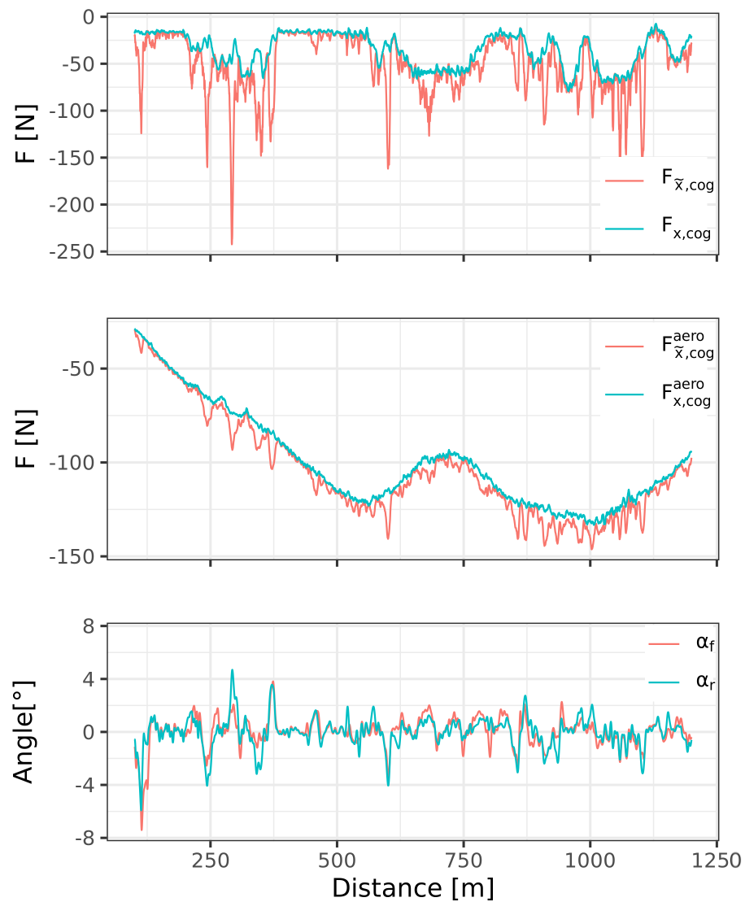}
\caption{Exemplary overview in the time domain of the forces acting against the direction of movement $F_\mathrm{\tilde{x},cog}$ compared to the  force in x-direction in the body coordinate system $F_\mathrm{x,cog}$  which is considered to be a theoretical optimum. As a reference, the side slip angles are shown as well. }
\label{fig_dd_forces_time}
\end{figure}

\begin{figure}
\centering
\begin{subfigure}[c]{0.8\textwidth}
\includegraphics[width=0.95\textwidth]{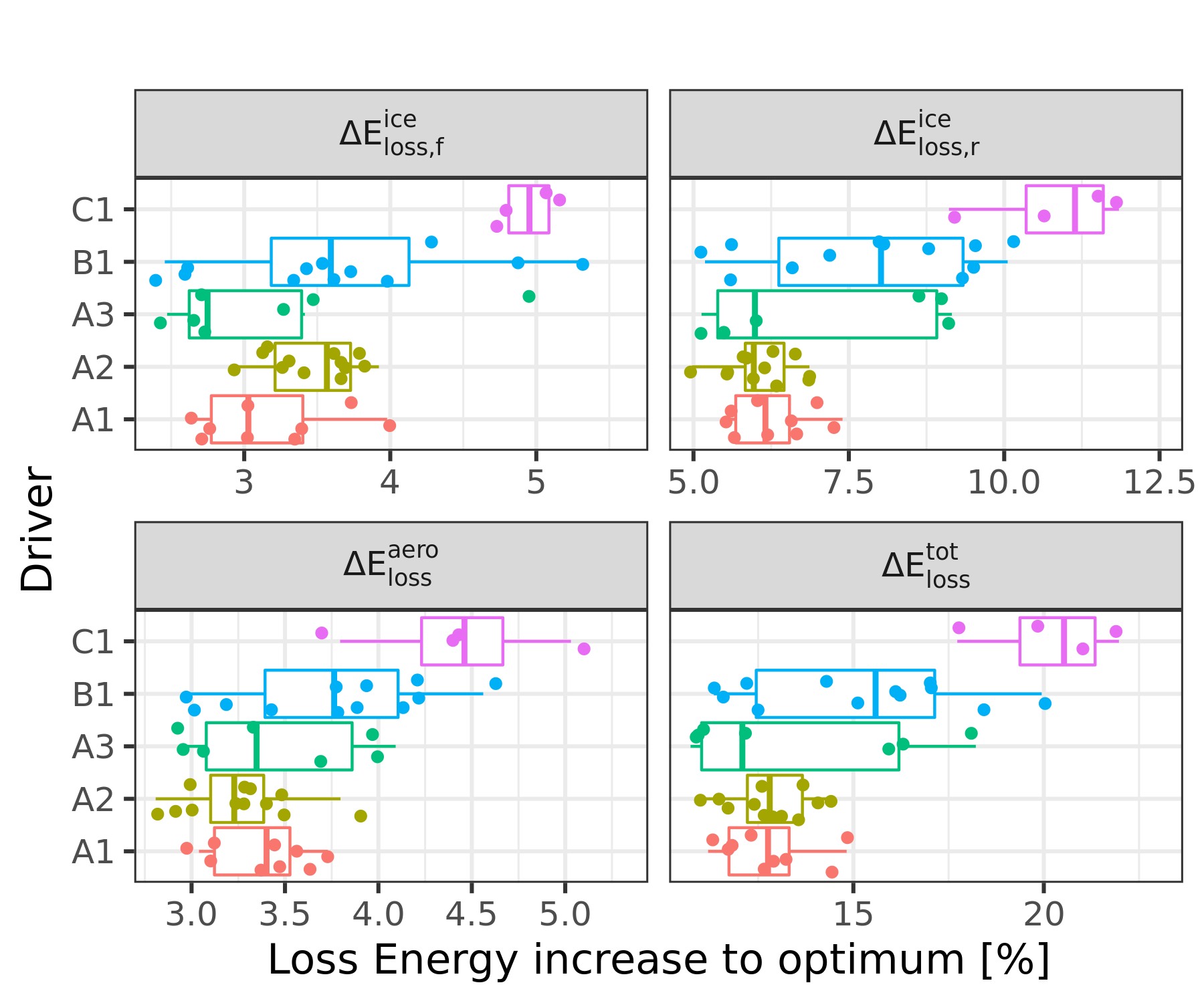}
\subcaption{Driver comparison on the track Königssee (KOE) }
\end{subfigure}
\begin{subfigure}[c]{0.8\textwidth}
\includegraphics[width=0.95\textwidth]{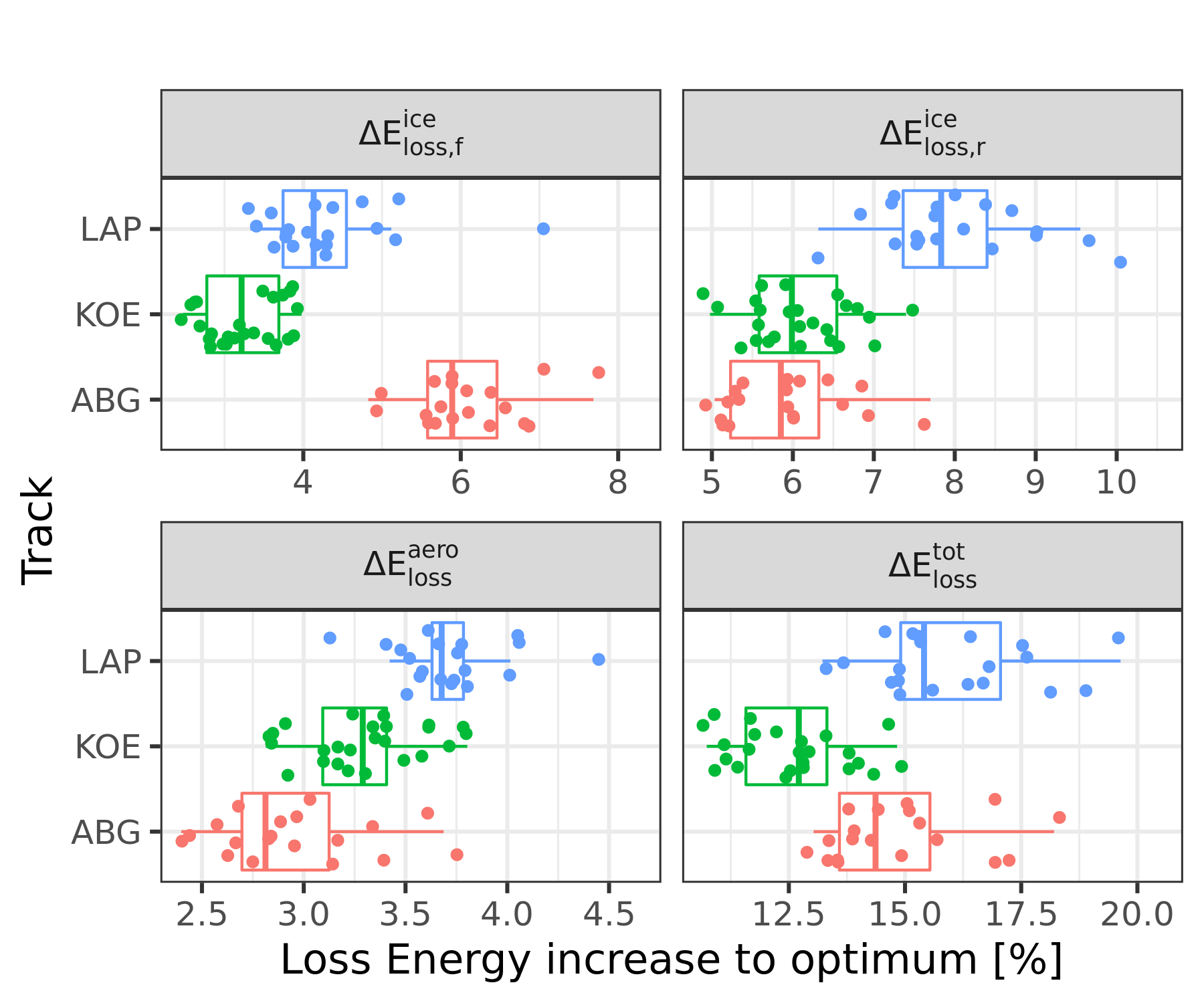}
\subcaption{Track comparison using only data from the top drivers (A1, A2, A3).}
\end{subfigure}
\caption{Energy comparisons for the different drivers and tracks. The relative increase of the loss energy with the ideal case as a basis is shown. It can be seen how much energy is lost at the front axle, rear axle, aerodynamically and overall. Every data point accounts for a complete run.}
\label{fig_energy_comp}
\end{figure}

We compare five drivers on the track in Königssee using the data set described in \autoref{tab:track_data}. The drivers are divided into the three classes A, B and C according to their level of experience. There are three top level drivers in the A class (the highest level of experience), one advanced driver in the B class (second highest level of experience) and one junior driver in the C class (lowest experience). 

At first we analyze the side slip angles and the steering angles between the drivers. \autoref{fig_boxplot_overview}a depicts the distributions of these variables in the form of boxplots. Since the values are small most of the time, the amount of rarer higher angles cannot be seen. Therefore, \autoref{fig_boxplot_overview}b shows the ratio of high absolute angles, i.e. over $2^\circ$ and $4^\circ$, respectively. \\
Looking at the steering angle $\delta$, Driver C1 steers distinctly less than all the other drivers, which holds also for high angle steering. There is not a single instance where C1 steered more than $4^\circ$. The low steering angles do not result in low slip angles at the front $\alpha_\mathrm{f}$ and the slip angles at the rear $\alpha_\mathrm{r}$ are the highest of all drivers. 
 Comparing the A-Drivers, A2 has the highest percentage of steering angles over $2^\circ$ and $4^\circ$ followed by A3 and A1. The slip angles at the front axle produce the same results. On the rear axle the differences are smaller, but here A1 has the lowest amount of high slip angles followed by A1 and A3. The drivers know from experience that drifts at the rear axle cause higher energy loss than at the front axle due to its shape (bigger length and different cross section), which is supported by our runner models (Section \ref{sec_runner_model}): The generated forces are distinctly higher than at the front axle for  given slip angles. However, by analyzing only the slip angles it cannot be determined which driving style is more advantageous.   \\
To solve this task we utilize the validated friction model to determine the forces which act against the driving direction  and calculate the relative loss energy increase as described in Section \ref{sec:driver_method}. \autoref{fig_dd_forces_time} depicts an exemplary overview of the actual and ideal forces at the center of gravity. It is visible how the side slip angles at front and rear axle lead to higher loss forces. \autoref{fig_energy_comp} shows the resulting distribution of loss energy increase for each driver and also for different tracks. \\

\subsubsection{Discussion of driving styles}
According to \autoref{fig_energy_comp}a, the A drivers have the lowest total losses, followed by  B1 and C1 when looking at the medians. This result is in close agreement with the defined classes. Driver C1 is the worst in all categories, therefore it is apparent that the driving style is disadvantageous and the low steering magnitude probably not deliberately chosen but a sign of uncertainty, as one would expect to see with less experienced drivers. Drivers A3 and B1 have a larger spread of values, indicating a more inconsistent driving style. Both have very good and bad runs. The different driving styles of driver A1 and A2 are interesting. We observe that A2 has distinctly higher losses at the front axle due to more steering but slightly less losses on the rear axle and due to aerodynamics. Overall, both approaches result in very similar overall losses. Consequently, in this case both driving styles are equally competitive. We suggest comparing these driving styles on different tracks in a future experiment. 
For a more detailed analysis, all the signals could be compared in the time or distance domain to  give the drivers clear instructions on how to achieve a different driving style if wanted.

\subsubsection{Track comparison}
As a side result from this work, \autoref{fig_energy_comp}b depicts the resulting energy differences between the tracks in La Plagne (LAP), Königssee (KOE), and Altenberg (ABG). For Altenberg and La Plagne only data from one top driver was available, for Königssee we used the data of the three A-drivers. Interestingly, the overall loss energy pieces together quite differently. Altenberg has distinctly higher losses at the front and less at the rear and aerodynamically. In comparison La Plagne has fewer losses at the front an more at rear. Königssee has overall the lowest losses. In summary, the tracks differ significantly from each other. We think that this information can help to improve the specific bobsled setup for these tracks.  

\section{Conclusion}
This work presents a data driven method to model the relevant aspects of ice friction for vehicle dynamics of a bobsled. The method is not specific to the sport of bobsleigh and can be transferred to other related areas. 
It shows that the developed friction model is suited for the usage in a driving simulator and capable of distinguishing the driving style differences of top drivers, which is a valuable asset for bobsleigh driver training. 
Since the driver evaluation depends only indirectly on outer circumstances, long-term driver performance analysis over several seasons could be conducted, which to the best of our knowledge has not been possible up to now. For future research we propose experiments to improve the lateral friction model for lower normal forces. For example ice house experiments where steering maneuvers with a bobsled (e.g. step steer) are conducted can  deliver cleaner data than track experiments because of a  distinctly lower amount of vibrations on the flat ice house surface. Furthermore, it should be investigated how other factors influence bobsled driving dynamics, for example the gliding velocity or ice surface properties such as temperature, roughness and hardness.

 \section*{Acknowledgments}
We thank the Bob- und Schlittenverband für Deutschland e.V. and its athletes for providing the bobsleigh equipment and conducting all test drives. Furthermore, we want to acknowledge ixent GmbH for fruitful discussions and providing  bobsled parameters. We also thank FES (Institut für Forschung und Entwicklung von Sportgeräten) for providing additional measurement data. 
Special thanks go to Sarah Lederer for proofreading the article. This work was generously supported by BMW AG.
 

\clearpage
\bibliography{jvs_bib}

\end{document}